\documentclass[12pt]{article}
\usepackage{graphics,epsfig}
\newcommand{\bce}{\begin{center}}
\newcommand{\ece}{\end{center}}
\newcommand{\beq}{\begin{equation}}
\newcommand{\eeq}{\end{equation}}
\newcommand{\bea}{\vspace{0.25cm}\begin{eqnarray}}
\newcommand{\eea}{\end{eqnarray}}

\newcommand{\ba}{\begin{array}}
\newcommand{\ea}{\end{array}}

\newcommand{\doublespace}{
    \renewcommand{\baselinestretch}{1.6}\large\normalsize}

\def\lsim{\mathrel{\rlap{\lower4pt\hbox{\hskip1pt$\sim$}}
    \raise1pt\hbox{$<$}}}     
\def\gsim{\mathrel{\rlap{\lower4pt\hbox{\hskip1pt$\sim$}}
    \raise1pt\hbox{$>$}}}     

\def\lsim{\mathrel{\rlap{\lower4pt\hbox{\hskip1pt$\sim$}}
    \raise1pt\hbox{$<$}}}         
\def\gsim{\mathrel{\rlap{\lower4pt\hbox{\hskip1pt$\sim$}}
    \raise1pt\hbox{$>$}}}         

\def\beq{\begin{equation}}
\def\endeq{\end{equation}}
\def\arr{\begin{eqnarray}}
\def\endarr{\end{eqnarray}}
\def\arrn{\begin{eqnarray*}}
\def\endarrn{\end{eqnarray*}}

\makeindex
\begin{document}
\vspace{2cm}

\begin{center}
{\bf \huge Annihilation of antiprotons on nuclear targets 
at low energies: a simple 
analysis.\\}
\vspace{1cm}
{\bf A. Bianconi$^{1,2)}$, G. Bonomi$^{1,2)}$
E. Lodi Rizzini$^{1,2)}$,
L. Venturelli$^{1,2)}$, A. Zenoni$^{1,2}$ } \medskip\\
{\small \sl
$^{1)}$Dipartimento di Chimica e Fisica per l'Ingegneria e 
per i Materiali, Universit\`a di Brescia, 
via Valotti 9, 25123 Brescia, Italy \\
$^{2)}$Istituto Nazionale di Fisica Nucleare,
Sezione di Pavia, Pavia, Italy \\}
\vspace{1cm}
{\bf \LARGE A b s t r a c t \bigskip\\}
\end{center}

We set up a plane wave impulse approximation 
(PWIA) formalism 
for the analysis of the  
annihilation cross sections 
of antinucleons on nuclear targets at very low momenta 
(below 100 MeV/c), where semiclassical approximations 
can't be applied. Since, as we test here,  
PWIA fails in reproducing 
the unexpected ``inversion'' behavior of the 
$\bar{p}p$ and $\bar{p}-$nucleus 
annihilation cross sections 
found in a recent experiment\cite{obe1,obe2} 
we discuss some further possibilities, with a  
special attention to the optical potential model. 

\medskip

\newpage
\doublespace

\section{Introduction}

Recents measurements\cite{obe1,obe2} of $\bar{p}$ annihilation on 
proton, deuteron and $^4$He targets show an interesting 
``inversion'' pattern at projectile momenta below 60 MeV/c: 
total annihilation cross 
sections on Deuteron and $^4$He are smaller than 
the same cross sections 
on free proton targets. Such a behavior is not seen at 
larger momenta, where total annihilation cross sections 
increase with the target mass number $A$, 
accordingly to intuition. There are no low energy 
data on larger nuclei, to clarify whether 
the decrease of cross sections with $A$ is a systematic 
phenomenon or specifical of a few light nuclei. 
In the case of $^4$He target the phenomenon is particularly 
impressing if one takes into account that the 
Coulomb attraction introduces a further factor $\sim$ $Z$
in any model for $\bar{p}-$nucleus annihilation, so a naive 
extimation of the ratio between low-energy 
$\bar{p}p$ and $\bar{p}^4$He 
annihilation cross section could be as small as 
$1/AZ$ $=$ 1/8. On the contrary, below 60 MeV/c the 
$\bar{p}-$D and $\bar{p}-^4$He annihilation cross sections 
are clearly smaller than $\bar{p}-p$ corresponding cross 
section. So, with $^4$He targets the most naive expectation 
largely overestimates the detected annihilation rate. 
A confirmation of these measurements comes from 
the recent data\cite{wid} on the widths and shifts 
of the electromagnetic 
levels of the antiprotonic deuterium. As demonstrated 
long ago\cite{tru} there is a relation between the 
width of a level in an antiprotonic atom and the 
$\bar{p}$ low energy annihilation 
rate on the corresponding nucleus. Although affected 
by large error bars, the comparison of the data on 
antiprotonic Hydrogen and Deuterium confirms the ``inversion'' 
behavior. For these last 
measurements a successful prediction 
existed\cite{wic} (see below for more details). 
In view of the possibility of a systematical analysis 
of $\bar{p}$ annihilation on complex  nuclei at very low 
energies 
at the $\bar{p}$ accelerator AD at CERN\cite{ad},
it is desirable to look for an understanding of what could 
be the properties of the interactions between $\bar{p}$ 
and nuclear matter, and to try to organize some model 
or framework aimed at this analysis. This will be 
the aim of this work, with special enphasis 
on the possible explainations of the inversion behavior. 

A simple 
black sphere potential model\cite{pro1} shows that 
the general properties of low energy scattering on strongly 
absorbing optical potentials don't contraddict the 
idea of ``inversion'': 
the S-wave contribution to the reaction cross section 
on a spherical imaginary potential ($V(r)$ $=$ $iW$ for 
$r$ $<$ $R$, and zero outside) is an increasing function of 
the strength parameter $W$ only at small $W$. When 
$W$ $\rightarrow$ $\infty$ the S-wave reaction cross section 
tends to zero. However, we have played a little with this 
model, and we have seen that: (1) If the strength $W$ is 
kept constant and $R$ increased (which seems 
representative of a comparison between different nuclear 
targets) the cross section tends to a constant plateau 
(after an initial increase) instead of decreasing. 
(2) If Coulomb interactions are included, the constant 
plateau is substituted by a $\sim Z$ $\sim$ $R^3$  
dependence. 
(3) The detected inversion behavior is present in a region 
where P-wave contributions are still large, perhaps 
dominant on nuclear targets, 
and where absorbing potentials with  
realistic shapes (we have tried 
Woods-Saxon reproducing known nuclear density properties)
fail in reproducing the inversion.   
(4) Recently we have been informed\cite{pro2} of a 3-body 
Faddeev-style analysis of the $\bar{p}-D$ annihilation 
problem, where inversion results only when 
additional physics (of a kind which is 
not contained in the black sphere model) is included. 
So, although a simple 
optical potential model presents intrinsic features 
which depress nuclear cross sections, it does 
not give yet a satisfactory explaination of what we see.

A specifical model for nuclear interactions, in 
this peculiar low energy domain, is still missing. 
From a fundamental point 
of view, it is not clear yet how much one can trust 
the concept of optical potential for describing 
hadronic annihilations (for a review about the  
theoretical developments in $\bar{p}-p$ annihilation 
see e.g. ref.\cite{dov1} and references therein). 
But even if one decides to trust it, it is desirable to 
have some physical connections between the nuclear potential 
and the underlying $\bar{p}-$nucleon interactions, not to 
reduce the physics of the problem just to a fit of 
radius, strength and a few more parameters. 
The traditional models used in nuclear physics 
for giving a meaning to the optical potential\cite{fesh} 
can hardly be applied here: 
they are based on the idea of 
a soft global rearrangement reaction 
proceeding through a set of $narrow$ excited states
of the full compound ``nucleus $+$ projectile''. 
This point will be discussed in the last part of this  
work, but presently there are no signals
of compound nucleus mechanisms. They should be accompanied 
by narrow resonances in the elastic channel and by 
complex fragmentation of the set of $A-1$ nucleons 
which don't annihilate. Such processes should be present 
at any $\bar{p}$ momentum of magnitude within the 
Fermi momentum ($\approx$ 200 MeV) or smaller. 
Narrow resonances have never been seen. Although we 
haven't fragmentation data specific to the inversion 
region, at $\bar{p}$ momenta $\sim$ 100 MeV/c 
there is no evidence\cite{lear1} of a 
reaction mechanism involving, at leading order, more than 
one target nucleon at a time: the number of 
annihilation processes leading to a fragmented residual 
is small enough to be explained in terms of final state 
interactions. 
The situation is in, some respect, reminiscent of 
high energy hadron-nucleus interactions, where the 
hard process is supposed to go mainly through single 
or multiple 
interaction between the projectile and single nucleons,
and sudden 
residual fragmentation, when present, can be 
attributed to rescattering or to initial state 
(projectile independent) correlations between target 
nucleons. 
However, a simplifying feature of high energy 
processes is missing 
here: because of the long wavelength of the projectile, all 
the semiclassical approximations and intuition are lost. 
At large energies, short wavelengths allow for 
exploitation of the Glauber formalism\cite{glau}. 
The lowest $\bar{p}$ incident momentum at which data of 
$\bar{p}-$nucleus elastic and reaction cross sections 
were interpreted in a satisfactory way with a 
Glauber-style model was 600 MeV/c\cite{aldo}. 
At lower momenta (300 and 200 MeV/c) such a model is only able 
to reproduce gross features of the differential cross sections, 
although it still revealed effective in reproducing 
integrated cross sections. We are now interested in a 
kinematical region where the Glauber model is surely 
not reliable anymore, i.e. with projectile momentum starting 
from around 100 MeV/c and going 
down to 1 MeV/c. At the lower end of this 
range, nonresonant annihilation processes are 
S-wave dominated with  any nuclear target.  
At momenta around 200 MeV/c 
the S-wave contribution is not the most 
relevant one (with nuclear targets). 
The ``inversion'' pattern, which  
appears below 60 MeV/c, seems to 
be associated with the transition to the S-wave dominance 
region, where surely the Glauber model can't be applied. 

The problem will be matched from different sides. 
We start from the easiest possible framework, i.e. 
the Plane Wave Impulse Approximation (PWIA). In
the Impulse Approximation 
one sums coherently single interactions of the 
projectile with any one of the target nucleons
(with the remaining $A-1$ acting as spectators),  
multiple interaction 
effects are disregarded and there is no reciprocal disturbance 
between scattering events on different nucleons. 
``Plane wave'' means that no nuclear distortion of the 
waves participating to the elementary scattering event 
$\bar{p}$ $+$ $p$ (or $\bar{p}$ $+$ $n$) 
$\rightarrow$ ``some final state'' 
is considered. At some extent, we include 
Coulomb correction effects 
(a more systematical analysis 
of this point is still going on and will be 
presented elsewhere). These corrections are very relevant 
and lead to differences in magnitude between, e.g., 
$\bar{p}-$D and $\bar{n}-$D cross sections when the 
projectile momentum is much below 100 MeV/c. 
Coulomb effects even produce curious phenomena like
a large difference between $\bar{p}-n$ cross sections 
in the two cases where the neutron target is $free$ or 
bound to a deuteron. 
For simplicity, 
here we will not take care of spin and isospin effects.   
Presently our knowledge on the 
elementary annihilation processes does not make a
discrimination among different spin and 
isospin channels very useful, although 
perhaps some information on the role of isospin 
is already available\cite{arm} or will be soon\cite{feli}. 
Waiting for systematic data on this point, 
in the following we will postulate approximative 
equality between $\bar{p}-p$ and $\bar{p}-n$ 
cross sections, apart for Coulomb effects. Despite
all these approximations the IA formalism, as is evident 
in the following, is rather complicate.  

Once Coulomb corrections are introduced, 
the most important limitation of the IA analysis is 
the lack of consideration for multiple interaction 
effects. So they will be discussed, at a qualitative level. 
We anticipate here that 
there are already some indications of the importance of 
these processes: (1) At larger energies, not to take 
rescattering into 
account leads to a wrong $A$ dependence of the annihilation 
rates, instead of the correct $A^{2/3}$ one. (2) At low 
energies, a  
Born leading order perturbative solution of the black sphere 
model does not produce inversion 
(at some extent the Born leading order, in a 
problem where the nucleus is approximated by a continuous 
potential, can be associated with discarding  
multiple scattering events in the realistic nuclear 
problem).  
(3) The strongest support 
to the role of multiple scattering in 
very low energy $\bar{p}-$nucleus physics 
comes from an analysis\cite{wic} of the related problem 
of the 
transition widths in antiprotonic deuterium. There, 
terms over IA have been studied and found relevant. 
Double scattering terms are fundamental in 
decreasing the IA effects, and the results of that 
work are in agreement with the recent data\cite{wid}.
(4) As evident in the following, IA does not lead to 
any inversion. 

The fact that IA does not lead to inversion can perhaps  
be expected in advance: a sum of amplitudes will not 
be smaller than any of them, if they all combine with 
the same phase (taking into 
account that interference effects due to the different 
space positions of the scattering centers 
are usually lost in the total cross sections). 
Since we have not enough information on the relative sign 
of $\bar{p}-p$ and $\bar{p}-n$ interactions, 
assuming them as equal depresses our chances of producing 
any inversion from the very beginning.  

Some interesting effects  
can be studied within the IA formalism.   
First, the IA 
allows one to calculate the Fermi motion smearing 
of any irregularity contained in the cross section for 
the interaction between any projectile and a single 
nucleon. This phenomenon is well known in many branches 
of nuclear physics and is simply due to the fact that 
bound nucleons move with random momentum 
$\sim$ 200 MeV/c. So, a specifical 
phenomenon (e.g. a resonance peak) which is present 
in the $\bar{p}-p$ cross section at a well defined 
value of the projectile momentum is spread through 
a kinematical region of range $\Delta p$ $\sim$ 
$p_{Fermi}$. We will show how this mechanism 
can produce an inversion behavior, and which 
conditions should be realized. According to the 
discussion accompanying fig.5 it will be evident that 
it is difficult, although not impossible, to support  
this explaination for the inversion. 

Another interesting point examined here is the 
sensitivity of the IA results to the high momentum components 
of the nuclear momentum distribution. In the following it 
is evident that 
IA results are aligned with the intuitive expectations 
at the condition that the nuclear momentum distribution 
is dominated by those momenta which are typical of 
mean field theories ($\sim$ 100 MeV/c). 
In the event of a certain consistence 
of the large momentum tail, the 
IA contribution to the annihilation rate is depressed 
by a small but observable factor $\sim$ 10\%). 
This effect is evidently not sufficient to justify inversion, 
but it can be 
interesting in itself, since it suggests that the 
annihilation rates are sensitive to 
those elusive contributions to the nuclear 
momentum distribution which come from direct 
nucleon-nucleon interactions. 

Finally, the last part of this work will be devoted 
to the study of the optical model in its simplest form, 
i.e. the black sphere. 
We are pushed by the fact that, as anticipated,  
the optical potential model does predict nuclear 
cross sections which are much smaller that naive 
geometrical expectations, with inversion or saturation 
behaviors. Actually, although first 
intuition suggests absorption cross sections 
$\sim$ $R^3$ or $R^2$ for any strong absorber of 
radius $R$, the general properties of low energy 
scattering 
suggest an $R^2$ law for the elastic 
total cross sections only, while 
esoenergetic reactions should 
roughly follow a behavior $\sigma$ 
$\sim$ $R/k$\cite{ll1}. The factor $1/k$ is a well known 
Bethe's prediction (which can be motivated by the 
idea that the probability for a spontaneous reaction is 
proportional to the time that any particle spends 
inside the interaction region), and $R$ can be put there 
to give the right dimension to a cross section 
(as we see later, one can motivate this 
expectance in terms of the scattering length). 
The $R/k$ law roughly works for 
any ``moderately reacting sphere''. The 
reaction properties of such a target 
can be expressed by an optical 
potential of the form $V$ $=$ $U$ $-$ $iW$.  
The imaginary part produces absorption of projectile flux 
according to the time dependence of the wavefunction 
$exp(-iEt)$ (since inside the potential region 
$E$ $=$ $T$ $+$ $V$, where $T$ is the initial kinetic 
energy of the projectile). The surprise is that 
with large values of $W$ (uncommon in traditional 
nuclear physics, but perhaps suitable to describe the 
violent annihilation phenomenon) low energy reaction 
cross sections predicted by the optical model 
become much smaller than $R/k$, and show no $R$ 
dependence. We will analyze in detail the behavior of the 
wavefunction in the ``strong optical model'', 
to understand how much of its predictions 
is related to the use of an optical potential and 
how much is general. The 
general conclusion will be that any 
model able to produce a ``vacuum region'' 
in the projectile flux will lead to a small reaction 
cross section, at the condition that the vacuum 
region surface diffuseness is small with respect to the 
vacuum region radius. 

The work is organized as follows: 
In the next section the PWIA formalism will be 
described, leading to eq.(\ref{free11}) where a nuclear cross 
section for a specific transition 
is written in terms of the corresponding $\bar{p}-$nucleon 
cross sections. This equation is actually more suited for    
antineutron projectiles, while antiproton interactions 
at low energy are affected by strong Coulomb effects. These 
effects 
are discussed in the third section, where eq.(\ref{free11})
is substituted by eq.(\ref{free12}). In the fourth section we 
discuss the main applications. First, we easily restore the high 
energy limit. Then, within certain approximations we introduce 
a low energy limit of total annihilation cross sections, 
in eq.(\ref{free17}). This equation expresses 
the ratio $\equiv Z R_A$ between the nuclear and single hadron 
annihilation cross sections in terms of the nuclear momentum 
distribution only (averaging on the details of the elementary 
$\bar{p}p$ and $\bar{p}n$ annihilations). Despite its 
semplicity, its calculation requires a many particle 
final state integration, so we limit ourselves to studying 
the simple case were two pions only are produced in the final 
state, presenting in fig.4 the results of this calculation
(which more or less confirm a priori expectations). 
For the nuclear momentum distribution, we adopt two 
simple mean-field examples. The last application of IA 
that we consider is the nuclear smearing of a resonance which 
could be found in the $\bar{p}p$ annihilation. 
In the fifth section we discuss qualitatively two 
effects which should be the subject of much more 
difficult works: the role of large momentum components of the 
nuclear distribution, and multiple scattering effects. 
In the further section we analyze the black sphere 
potential model, compare the two limiting cases 
of high and low energy projectiles, generalize some results 
and discuss their possible relation with the 
detected inversion data. 
The last section is just 
a short recollection of the main points. 

\section{General PWIA framework}

We start from the amplitude 
$T_{A,n}(\vec p,\vec \pi_1,...\vec \pi_N)$ 
for the reaction $A(\bar{p},N\pi)(A-1)_n$, where the annihilation 
of a $\bar{p}$ with momentum $\vec p$ takes place on a 
single nucleon inside a nucleus 
leaving the residual sustem of $A-1$ nucleons in a state $n$, 
and producing a number $N$ of final state particles 
with momenta $\vec \pi_1,...,\vec \pi_N$ 
(Pions and, in smaller number,  
Kaons, but in the following we will call them 
all ``pions'', and the target nucleon will be called ``proton''). 
We know from experimental evidence\cite{af1} 
that the largest part of 
annihilation events imply production of up to  
seven pions).  The elementary 
process $\bar{p} + p \rightarrow N\pi$ 
(or $\bar{p} + n \rightarrow N\pi$)  
is described by the $(N+2)-$point amplitude 
$T(\vec r, \vec r_1, \vec r_3, ... \vec r_{N+2})$, 
where $\vec r$ refers to $\bar{p}$, $\vec r_1$ to the 
annihilated target nucleon, and $\vec r_3, ... r_{N+2}$ 
to the produced pions. The ``missing'' radius $\vec r_2$
will be used to describe the position of the center of 
mass of the residual $A-1$ nucleons, and $\vec x_2$
will represent the set of $3(A-1)$ 
coordinates of the residual $A-1$ nucleons with 
respect to their center of mass $\vec r_2$ (only $3(A-2)$ 
among these coordinates are independent). 

\arrn
T_{A,n}(\vec p,\vec \pi_1,...\vec \pi_N)\ =\  
\int 
e^{-i\sum_j\pi_j r_j} [\Psi^*_{A-1}(r_2,x_2)]_n
T(r,r_1,r_j) 
e^{ipr} \Psi_A(r_1,r_2,x_2)
\endarrn
\arr
d^3r d^3r_1 d^3r_2 d^{3(A-2)}x_2 d^{3N}r_j, 
\ \ (j = 3,..N+2). 
\label{tan}
\endarr

In particular:  

\beq
\Psi_A(r_1,r_2,x_2)\ \equiv\ exp(iPX)\Phi_A(x_1,x_2),\ 
\ \Psi_{A-1}(r_2,x_2)\ \equiv\ exp(iP_2 r_2)\Phi_{A-1,n}(x_2),
\label{psi}
\endeq
where $\vec P$ and $\vec P_2$ are the momenta of the initial 
nucleus and of the final residual nuclear system, and $\vec X$ 
is the position of the center of mass of the initial nucleus. 

Rigorously, in the previous matrix element Coulomb 
functions should appear, otherwise 
we are considering $\bar{n}$ 
scattering. Below we 
discuss the consequences implied when Coulomb interactions are 
properly taken into account. 

In the calculations we will rely on the closure approximation 
on the internal states of the (unobserved) residual.  
The cross section $\sigma(\vec p,\vec \pi_j)$,
corresponding to 
a given choice of $\vec p, \vec \pi_1,...\vec \pi_N$, 
is proportional to $\sum_n \vert T_{A,n} \vert^2$:

\beq
\sum_n \vert T_{A,n}(p,\pi_1,...\pi_N)\vert^2\ =\  
\sum_n \int \int dx_2dx_2' \Phi_{A-1,n}(x_2')
\Phi^*_{A-1,n}(x_2)......
\endeq
If the sum is on all the possible internal 
states of the residual, then:

\beq
\sum_n\Phi_{A-1,n}(x_2')
\Phi^*_{A-1,n}(x_2)\ =\ \delta^{3(A-2)}
(\vec x_2 - \vec x_2').
\label{clos}
\endeq
Rigorously the closure approximation is never exact, however 
the available energy in the game is high enough for a large 
continuum of residual states being accessible. 
From a practical point of view applying the closure means 
that, in the following integrations, all variables $r$, $R_1$ 
etc. will have their primed counterpart $r'$, $R_1'$ etc.  
with the exception of the $3(A-1)$ coordinates $\vec x_2$
(of which $3(A-2)$ only are integration 
variables), and the residual $internal$ functions $\Phi_{A-1,n}$ 
disappear. Then:  

\arrn
\sigma(\vec p,\vec \pi_j)\ \propto\ 
\sum_n \vert T_{A,n}(p,\pi_1,...\pi_N)\vert^2\ =\  
\int 
e^{i\sum_j\pi_j(r_j'-r_j) + ip(r-r') + i(P_2'r_2' - P_2r_2)} 
\endarrn
\arr
T^*(r',r_1',r_j')T(r,r_1,r_j) 
\Psi^*_A(r_1',r_2',x_2)\Psi_A(r_1,r_2,x_2) 
dr dr' dr_1 dr_1' dr_2 dr_2' dx_2 d^{3N}r_j d^{3N}r_j'
\label{s1}
\endarr

We clearly need to express everything in terms of the 
elementary scattering amplitude 
with all particles of well defined 
momentum. 
To get to this we need to decompose the bound nucleon state 
into Fourier components:

\arrn
\int d^3r_2 d^3r_2' d^{3(A-2)}x_2 
e^{-iP_2r_2+iP_2'r_2'}
\Psi^*_A(r_1',r_2',x_2)\Psi_A(r_1,r_2,x_2) \ \equiv
\endarrn
\arr
\equiv\ 
{1 \over {(2\pi)^6}} \int \int d^3k d^3k' F(k,k')
e^{-ik'r_1'+ikr_1}
\label{ft1}
\endarr
that implies: 
\arrn
F(k,k')\ =\ \int \int d^3r_1 d^3r_1' e^{ik'r_1'-ikr_1}
\int d^3r_2 d^3r_2' d^{3(A-2)}x_2 
e^{-iP_2r_2+iP_2'r_2'}
\endarrn
\arr
\Psi_A^*(r_1',r_2',x_2)\Psi_A(r_1,r_2,x_2).
\label{ft2}
\endarr

Writing esplicitely the nuclear wavefunction, according
to eq.(\ref{psi}), and using the substitutions:
\beq
\vec r_1\ \equiv\ \vec r_2 + \vec x_1,\ 
\vec X\ \equiv\ {{\vec r_1 + (A-1)\vec r_2} \over A}\ 
=\ \vec r_2 + {{\vec x_1} \over A},
\label{coo}
\endeq 
we obtain:  

\arrn
F(k,k') \ =\ \int dr_1 dr_1' dr^2 d{r'}^2dx^2
\endarrn
\arrn
e^{-iP_2r_2 + iP(r_2+x_1/A)-ik(r_2+x_1)}
e^{iP_2'r_2' - iP(r_2'+x_1'/A)+ik'(r_2'+x_1')}
\Phi_A(x_1,x_2)\Phi_A^*(x_1',x_2)\ =
\endarrn
\arr 
=\ (2 \pi)^6 \delta^3(P_2+k-P) \delta^3(P_2'+k'-P)
S(k-P/A,k'-P/A),
\label{ft3}
\endarr

where 
\arrn
S(k,k')\ \equiv\ \int d^3x_1 d^3x_1' e^{-ik_1x_1+ik_1'x_1'}
\int d^{3(A-2)}x_2
\Phi_A(x_1,x_2)\Phi_A^*(x_1',x_2)
\ \equiv 
\endarrn
\arr
\equiv\ \int dE S(k,k',E).
\label{sf1}
\endarr

The function $S(k,k',E)$ is the nuclear spectral function 
for one nucleon removal\cite{bgp}. 
In a mean field approximation it 
has the known shell model form 
\beq
S(k,k',E)_{sm}\ =\ \sum_\alpha \delta(E-E_\alpha)
\phi_\alpha(\vec k)\phi^*_\alpha(\vec k').
\label{sf2}
\endeq
However, due to few$-$body direct interactions, 
it is well known that the continuum deep energy 
components of $S$ can play a strong role when large  
energy releases are involved in the elementary 
processes\cite{bgp}. 

In a treatment where initial and final
state waves are affected by nuclear medium distortion 
(Distorted Wave Impulse Approximation DWIA)
momentum 
exchanges affect initial and final states, and all the 
non$-$diagonal components of $S(k,k')$ are involved. 
However, in a $PWIA$ treatment the remaining part of the 
integral contains a function $\delta^3(k-k')$, as we shall 
see later. So we can write from now on: 

\arrn
F(k,k')\ \rightarrow\ 
(2 \pi)^6 \delta^3(P_2+k-P) \delta^3(P_2'-P_2)
S(k_P/A,k-P/A) 
\rightarrow
\endarrn
\arr
\rightarrow\ 
(2 \pi)^6 \delta^3(P_2+k) \delta^3(P_2'-P_2)
S(k,k).  
\label{ft4}
\endarr

The last equality follows from choosing the laboratory  
frame, where 
the target nucleus is at rest. Now we may rewrite: 

\arrn
\sigma(\vec p,\vec \pi_j)\ \propto\ 
\sum_n \vert T_{A,n}(p,\pi_1,...\pi_N)\vert^2\ =\ 
\int 
e^{i\sum_j\pi_j(r_j'-r_j) + ip(r-r') + i(P_2'r_2' - P_2r_2)} 
\endarrn
\arrn
T^*(r',r_1',r_j')T(r,r_1,r_j) 
\Psi^*_A(r_1',r_2',x_2)\Psi_A(r_1,r_2,x_2) 
dr dr' dr_1 dr_1' dr_2 dr_2' dx_2 dr_j dr_j'\ =
\endarrn
\arrn
=\ {1 \over {(2\pi)^6}}\int d^3kd^3k'
e^{i\sum_j\pi_j(r_j'-r_j) + ip(r-r')  
+ i(kr_1-k'r_1')}
\endarrn
\arr
T^*(r',r_1',r_j')T(r,r_1,r_j) 
F(k,k')
dr dr' dr_1 dr_1'.  
\label{s2}
\endarr

In the previous equation it is easy to recognize the 
amplitude for the elementary process in momentum 
representation: 

\beq
G(p,k,\pi_j)\ =\ \int d^3r d^3r_1 d^{3N}r_j
e^{-i\sum_j\pi_jr_j + ipr + ikr_1} T(r,r_1,r_j)
\label{t1}
\endeq

However it is obvious that $T$ will not depend on the 
overall position of the process in the space, so its
Fourier transform $G$ must 
contain a function $\delta^3(\sum_j\pi_j-p-k)$. To esplicitly 
show it, one must use relative coordinates. One possible 
choice is: 
$\vec \xi_\pi$ $=$ $\sum_j \vec r_j /N$ (center of mass of 
the outgoing pions),
$\vec\xi_j$ $=$ $\vec r_j-\vec \xi_\pi$, $\xi_p$ $=$ 
$(\vec r_1+\vec r)/2$ (center of mass of 
the colliding $\bar{p}$ and $p$), $\vec \xi_1$ $=$ 
$\vec r_1-\vec r_2$, and finally $\vec \xi$ $=$ 
$\vec \xi_\pi-\vec \xi_p$ and $\vec Y$ $=$ 
$\alpha \vec \xi_\pi + \beta \vec \xi_p$ 
($\alpha$ $=$ $N m_\pi/(Nm_\pi+2m_p)$, $\beta$ $=$ 
$1-\alpha$). So $\xi$ is the distance between the 
initial and final state centers of mass, and $Y$ the
``center of mass of the centers of mass''. 
Calculating $G$ with this new set of coordinates, 
one patiently arrives to: 

\beq
G(p,k,\pi_j)\ =\ (2\pi)^3\delta^3(p+k-\sum\pi_j) 
T(p+k,p-k,\pi_j),
\label{t2}
\endeq

\beq
T(p+k,p-k,\pi_j)\ \equiv\ 
\int d^3\xi d^3\xi_1 d^{3N-1}\xi_j
e^{-i(p+k)\xi-i(p-k)\xi_1/2-i\sum_j\pi_j\xi_j/N} 
T(\xi,\xi_1,\xi_j)
\label{t3}
\endeq

Although written in a more complicated form, the latter 
amplitude $T$ corresponds to the function $f(E,\theta)$ in the 
ordinary nonrelativistic scattering theory, where the 
problem is written with respect to relative coordinates, 
the target is given infinite mass, the projectile mass is 
changed to $m_1m_2/(m_1+m_2)$, and the wavefunction of the relative 
motion is written in the form $\exp(i\vec p\vec r)+
f(E,\theta)exp(ip'r)/r$. The amplitude $f$ is the energy 
conserving scattering amplitude in momentum representation. 

It is easier to see this if we start from eq.(\ref{s2})
and substitute the nuclear target 
with a free proton with momentum $\vec k_o$. 
This is equivalent to performing the substitution:  
\arr
F(k,k')\ \rightarrow\ (2\pi)^6 \delta^3(\vec k - \vec k_o)
\delta^3(\vec k' - \vec k_o)
\label{free1}
\endarr
in eq.(\ref{s2}), together with taking properly into 
account the change of the 
initial flux and of the final state phase space. 
To see this we rewrite 
eq.(\ref{s2}) in terms of the frequency of  
events $dW$ for a detected value of the recoil momentum 
of the residual $R$ (in the previous formalism closure has been 
applied on residual $internal$ states only) and detected values 
of all the produced mesons: 

\arrn
dW(\bar{p}+A\rightarrow N\pi+R)
\ =\ 
C d\Phi_{N+1} {1 \over {(2\pi)^6}}\int d^3kd^3k'
e^{i\sum_j\pi_j(r_j'-r_j) + ip(r-r')  
+ i(kr_1-k'r_1')}
\endarrn
\arr
T^*(r',r_1',r_j')T(r,r_1,r_j) 
F(k,k') dr dr' dr_1 dr_1' dr_j dr_j'
\label{free2}
\endarr
where $C$ is a constant and $d\Phi_{N+1}$ is the phase space for 
emission of $N+1$ particles with energy conservation (momentum 
conservation is already present in the integral via the $\delta$ 
functions in $F$ and $G$). 
When the substitution (\ref{free1}) and the change of phase space 
are performed, eq.(\ref{free2}) becomes:

\arr
dW(\bar{p}+p\rightarrow N\pi)
\ =\ C d\Phi_N
\vert G(p,k_o,\pi_j)\vert^2 
\label{free3}
\endarr
where $d\Phi_N$ is the phase space for emission of $N$ 
particles with energy conservation, and 
the constant $C$ is the same as in eq.(\ref{free2}). 
In the last equation (see eq.(\ref{t2})) 
a squared delta function is present: 
$[\delta^3(p+k_o-\sum\pi_j)]^2$. The usual way \cite{ll4} 
to treat this excess of singularity is by the substitution:  
\arr
(2\pi)^6[\delta^3(p+k_o-\sum\pi_j)]^2\ 
\rightarrow\ 
(2\pi)^3 V \delta^3(p+k_o-\sum\pi_j) 
\label{free4}
\endarr
where $V$ is an overall normalization volume. 
This volume will disappear from the final expression 
for the cross section, so that one can put $V$ $=$ 1 
from the very beginning. 
Taking into account eq.(\ref{t2}), we may rewrite 
eq.(\ref{free3}): 

\arr
dW(\bar{p}+p\rightarrow N\pi)
\ =\ C d\Phi_N
(2\pi)^3\delta^3(p+k_o-\sum\pi_j) 
\vert T(p+k_o,p-k_o,\pi_j)\vert^2.
\label{free5}
\endarr

Dividing by the incoming flux one gets the corresponding 
cross section, which of course is nonzero for conserved 
momentum and energy.

Exploiting the definition (\ref{t1}) of $G$, 
writing explicitely $F$ and $G$ 
according to eq.(\ref{ft4}) and eq.(\ref{t2})
and exploiting the two delta functions contained in $F$ 
for removing $k$ and $k'$ integrations, eq.(\ref{free2})
becomes:

\arrn
dW(\bar{p}+A\rightarrow N\pi+R)
\ = 
\endarrn
\arr
=\ C d\Phi(N+1) 
(2\pi)^6\bigg\{[\delta^3(p+k-\sum\pi_j)]^2 
\vert T(p+k,p-k,\pi_j)\vert^2\
S(k,k)\bigg\}_{k = -P_2}
\label{free6a}
\endarr

and once one of the two delta functions is removed 
(using the same 
trick as previously) this last equation becomes: 

\arrn
dW(\bar{p}+A\rightarrow N\pi+R)
\ =
\endarrn
\arr
=\ C d\Phi(N+1) 
(2\pi)^3\bigg\{\delta^3(p+k-\sum\pi_j) 
\vert T(p+k,p-k,\pi_j)\vert^2
S(k,k)\bigg\}_{k = -P_2}.
\label{free6}
\endarr

The same comments following eq.(\ref{free5}) can now be 
made about eq.(\ref{free6}). It is singular on the 
momentum conservation shell, 
because the global recoil momentum 
of the residual is supposedly detected. The constant $C$ 
is the same in eq.(\ref{free5}) and eq.(\ref{free6}) 
if the normalization of $S$ is chosen as:  
\arr
\int S(k,k) d^3k/(2\pi)^3 \ =\ A
\label{sfree}
\endarr
and this 
allows for comparison between nuclear and free proton 
cross sections. 

However, it is unlikely that nuclear 
recoil information are present in the experiment, so that 
we probably need to integrate the last equation on the 
phase space of the recoil center of mass 
$d^3k/(2\pi)^3$. The correct phase space element would 
require a factor $1/2E$ $\approx$ $1/2M_r$.  
However we must take into account that, in a 
relativistic treatment, spinors are normalized like 
$\bar{u} u$ $=$ $2M$, so that the net effect of the 
$1/2E$ factor disappears from the cross sections as 
far as we may neglect the kinetic energy of the residual 
center of mass (of course the phase space of the 
emitted mesons, considered in the following, 
will be relativistic). 
After this integration the cross section is not singular anymore 
on the 3-momentum conservation shell. It is still 
singular on the energy shell. The form of the energy conservation 
will be simplified by neglecting the 
energy of the nuclear recoil. 
We have:  

\arr
dW(\bar{p}+A\rightarrow N\pi+X)
\ =\ 
C d\Phi_N 
(2\pi)^3\bigg\{
\vert T(p+k,p-k,\pi_j)\vert^2
S(k,k)\bigg\}_{\vec k = \sum \vec \pi_j - \vec p}
\label{free7}
\endarr

To get to cross sections we still have to consider flux 
factors. This point is delicate: at the small momenta of our 
interest, the 
most evident behavior of the cross sections is dictated by 
the flux factor $1/p$. For collisions between two 
particles with masses and 4-momenta 
$m_1,m_2,p^\mu,P^\mu$, the relation 
between $d\sigma$ and the above used $dW$ is, neglecting 
spins, $d\sigma$ $=$ $dWE_1E_2/I$ ($E_i$ in relativistic sense) 
where $I$ $\equiv$ 
$\sqrt{(p_\mu P^\mu)^2 - m_1^2m_2^2}$ (see e.g. \cite{ll4}, taking 
into account that notations for $dW$ are not exactly the same). 
When the second particle is at rest (laboratory frame) 
the flux factor $I/E_1E_2$ coincides exactly with 
the projectile velocity 
$\beta_1$ in the laboratory frame. When the target 
is not at rest, but moves with nonrelativistic velocity, 
the flux factor coincides with the relative velocity
up to relativistic corrections.   
So, neglecting spin factors, we can write the two comparable 
equations: 

\arr
d\sigma(\bar{p}+p\rightarrow N\pi)
\ =\ {C \over \beta_{rel}} d\Phi_N
(2\pi)^3\delta^3(p+k-\sum\pi_j) 
\vert T(p+k,p-k,\pi_j)\vert^2
\label{free8}
\endarr
for the free proton target with momentum $k$ 
($\beta_{rel}$ becomes simply $\beta$ when $\vec k_o$ $=$ 0), 
and 

\arr
d\sigma(\bar{p}+A\rightarrow N\pi+X)
\ =\ 
{C \over \beta} d\Phi_N 
(2\pi)^3\bigg\{
\vert T(p+k,p-k,\pi_j)\vert^2
S(k,k)\bigg\}_{\vec k = \sum \vec \pi_j - \vec p}
\label{free9}
\endarr
for a nuclear target at rest 
without detection of the residual. 
Writing: 

\arr
S(k,k)_{\vec k = \sum \vec \pi_j - \vec p}\ =\ 
\int {{d^3k}\over{(2\pi)^3}} S(k,k) 
(2\pi)^3\delta^3(\vec p+\vec k-\sum\vec \pi)
\label{free10}
\endarr
and substituting this equation in eq.(\ref{free9}) we 
get the relation between the two cross sections:

\arr
d\sigma_{A,N}(\vec p,0,\vec \pi_j)
\ =\ \int {{d^3k}\over {(2\pi)^3}} 
S(k,k) {{\beta(\vec p,\vec k)} \over {\beta(\vec p,0)}}
d\sigma_{1,N}(\vec p,\vec k,\vec \pi_j)
\label{free11}
\endarr
where: 
$d\sigma_{A,N}(\vec p,0,\vec \pi_j)$ is the cross section for 
annihilation of an antiproton with initial momentum $\vec p$ 
on a target nucleus at rest, with production of $N$ pions with 
detected momenta $\vec \pi_j$ in the phase space element 
$\delta(E_{fin}-E_{in})\prod d^3\pi_j/[2E_j(2\pi)^3]$,
and no information on the residual nucleus; 
$d\sigma_{1,N}(\vec p,\vec k,\vec \pi_j)$ is the cross section for 
annihilation between an antiproton with initial momentum $\vec p$ 
and a proton with momentum $\vec k$, 
with production of $N$ pions with 
detected momenta $\vec \pi_j$ in the same 
phase space element as before; 
$\beta(\vec p,\vec k)$ is 
the relative velocity between two incoming particles with 
momenta $\vec p$ and $\vec k$; $S(k,k)$ must be normalized 
so that $\int {{d^3k}\over {(2\pi)^3}} S(k,k)$ $=$ $A$. 
The nuclear cross section $d\sigma_{A,N}$ 
is not momentum conserving. 

The Coulomb corrected version of this equation (see below),
together with its low energy application,  
is the main result of the calculations contained in this paper. 
The most evident observation is that it contains an 
$uncoherent$ sum over the elementary scattering possibilities, 
i.e. the integral is over cross sections, not over 
amplitudes. It must be stressed that this lack of interference 
effects is due to the PWIA. In presence of rescattering 
we would find terms containing $S(k,k')$ with different 
$k$ and $k'$, and products $TT'$ of amplitudes corresponding 
to different processes. 

\section{Coulomb corrections}

The above equation (\ref{free11}) is correct, 
within PWIA, for describing interactions of an  
$\bar{n}$ 
with a nucleus. For a $\bar{p}$ projectile it 
requires further relevant corrections 
because of Coulomb effects\cite{cph}. 
Coulomb attraction, 
at semiclassical level, implies focusing of the incoming $\bar{p}$ 
flux towards the scattering center. For pure S-wave 
low-energy annihilations on pointlike targets 
this implies a further $Z/\beta$ factor 
(times a Z-independent constant that we reabsorb in 
the target charge $Z$ not to 
overload formulas) 
in $d\sigma_{1,N}$\cite{ll1}. This correction has 
been estimated since long from the ratio 
$\vert \Psi_c/\Psi_o\vert^2_{r = 0}$ between the 
coulomb problem solution $\Psi_c$ and the free motion 
wave $\Psi_o$. 

For the case of an extended target of nuclear size, 
with nonsingular charge distribution, we have 
numerically performed some test, representing the nucleus as 
a Wood-Saxon imaginary potential well, 
comparing the results of the two cases charged/neutral   
projectile. The charge distribution of the target is 
assumed homogeneous up to a radius $r_{charge}$ $=$ 
1.25 $A^{1/3}$ fm. 
The detailed and systematic presentation of 
these Coulomb corrections 
will be the subject of a forthcoming paper. 
Here we report a simple example, in figs.1,2 and 3, and 
anticipate some general results. 

\begin{figure}[htp]
\begin{center}
\mbox{
\epsfig{file=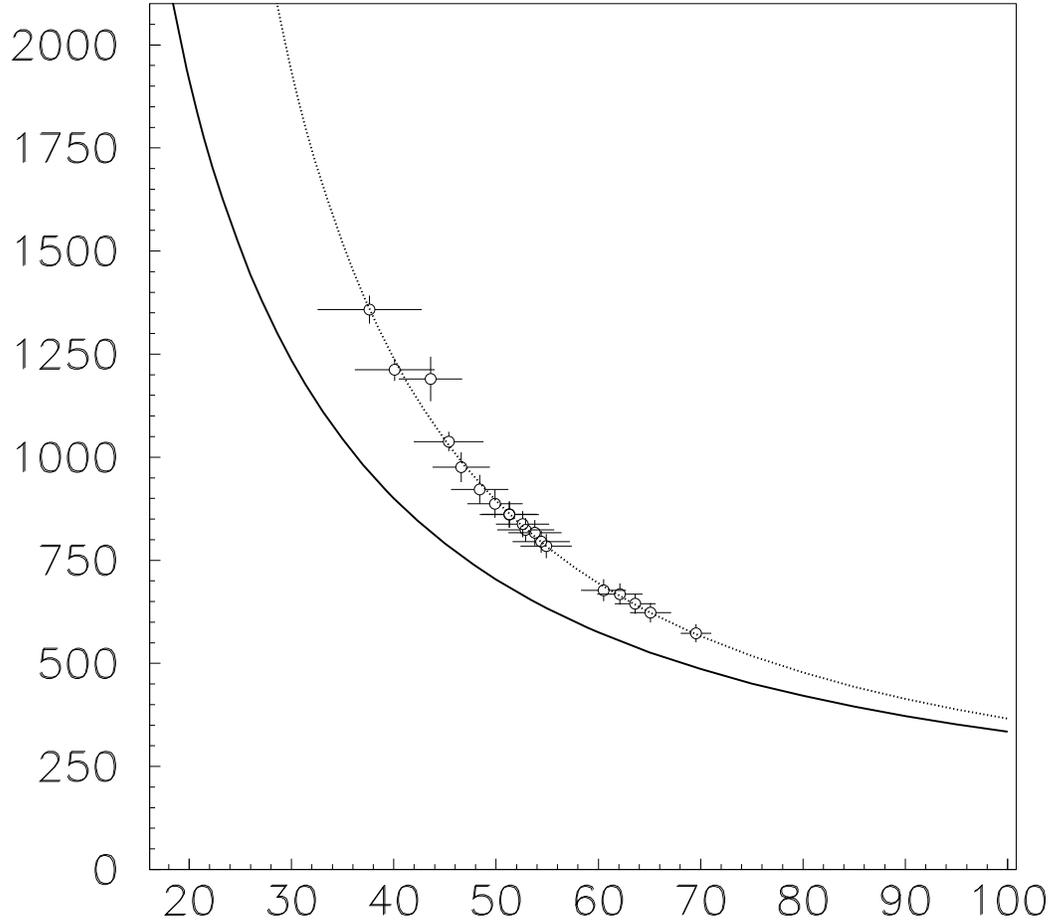,width=0.9\linewidth}}
\end{center}
\caption[]
{\small\it \label{fig1}
Continuous line: 
the cross section (mb) 
for annihilation of $\bar{p}$ on a proton 
target, simulated by 
an imaginary potential of Woods-Saxon kind, with radius 
1.3 fm, diffuseness 0.6 fm and strength 33 MeV, without 
inclusion of Coulomb interaction. Dotted line: the same, with 
Coulomb interaction included. The two curves are given as 
functions of the projectile momentum (MeV/c) in the laboratory
frame. Center of mass corrections are included in the calculation. 
The parameters of the potential are chosen so that the 
dotted curve fits reasonably the available 
data in this region (see text). These data, explicitely reported 
in the figure, are taken from references \cite{obe1}, 
\cite{obe3} and \cite{obe4}.}
\end{figure}

\begin{figure}[htp]
\begin{center}
\mbox{
\epsfig{file=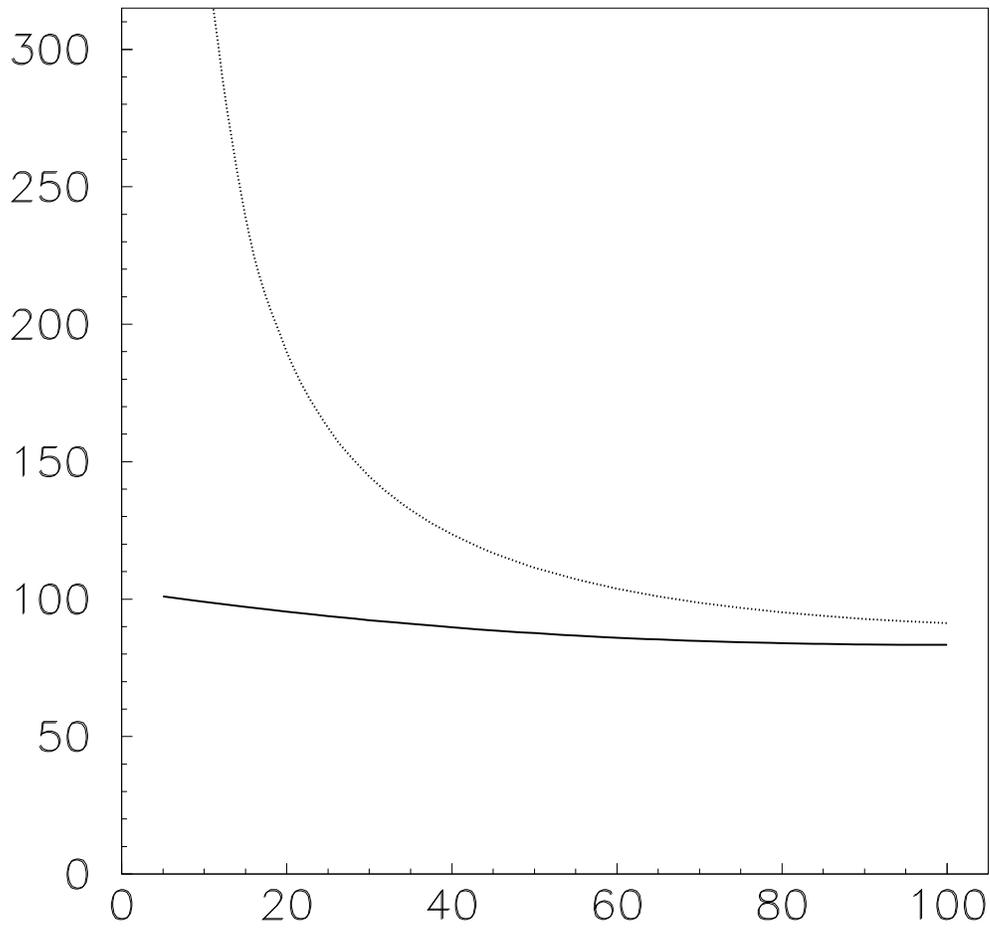,width=0.9\linewidth}}
\end{center}
\caption[]
{\small\it \label{fig2}
Continuous and dotted lines represent the same cross sections 
as in fig.1, but multiplied by $p$ (MeV/c), 
the projectile momentum in the laboratory. 
}
\end{figure}

\begin{figure}[htp]
\begin{center}
\mbox{
\epsfig{file=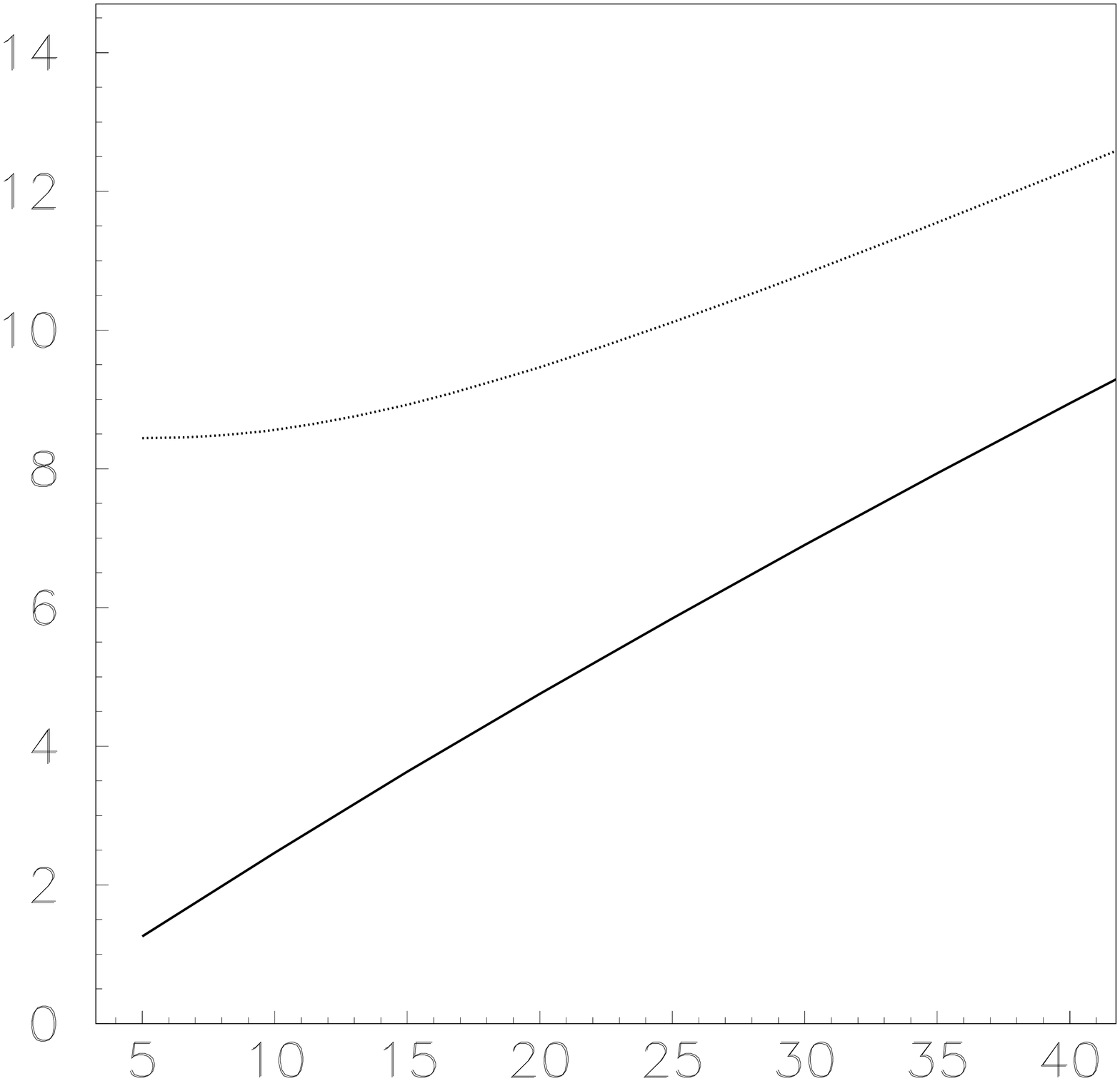,width=0.9\linewidth}}
\end{center}
\caption[]
{\small\it \label{fig3}
Continuous and dotted line represent the same cross sections 
as in fig.1, but multiplied by $p^2$. 
}
\end{figure}

The total cross section 
for $\bar{p}p$ annihilation can be well fitted 
in the range from 37 MeV/c to 70 MeV/c by an  
imaginary potential, of Woods-Saxon kind, with range  
1.3 fm, diffuseness 0.6 fm and strength 33 MeV. 
The consequent cross section is showed in fig.1, 
together with the ``uncharged'' cross section. 
Center of mass corrections have been taken into account 
in the calculation.   
In the next two figures the behavior of the 
quantities $p\sigma$ and 
$p^2\sigma$ is showed for both ``charged'' and 
``uncharged'' cross sections. In these nonrelativistic 
conditions momenta and velocities are proportional 
within a few percents. From the figures it is possible 
to see that both cross sections follow the $1/\beta$ 
standard law at the larger momenta of the considered range, 
but Coulomb deviations become more and more evident 
as the energy decreases, as already predicted\cite{cph}. 
The correcting factor becomes steadily proportional to  
$1/\beta$ for momenta 
below 10 MeV/c, as we can see by comparing  
figs. 2 and 3 at the left end of the scale. 
At larger momenta one can speak of 
a factor $F(\beta)$ some way between 
1 (for larger $p$) and $Z/\beta$ (for smaller $p$). 
We have even tried electron screened potentials, but 
without seeing any relevant screening 
effect down to $p$ $=$ 1 MeV/c (probably screening 
will cause a cut off of the Coulomb effect at some 
small $p$, but this does not seem to happen in the 
considered range, even for heavy nuclei). 
When the reaction cross section is decomposed into partial 
wave contributions, 
the correcting factor seems to affect the same way all 
partial waves (for $p$ $<$ 100 MeV/c, we find a contribution 
of some percent in the $D$ wave). 
Despite the very small involved momenta 
and angular momenta, this behavior 
allows for a semiclassical interpretation in terms 
of focusing of projectile trajectories.  
Similar conclusions remain valid even when 
we simulate much heavier nuclei. 

The previous Coulomb corrections have been calculated 
assuming the nucleus as an absorbing potential. However, 
because the Coulomb focusing 
occurs on a scale which is sensibly larger than the 
nuclear radius and at $\bar{p}$ momenta which are not much larger 
than the Fermi momentum, its effect should not depend on 
the Fermi motion of the target constituents: indeed, 
during the time needed by Coulomb forces to deflect the incoming 
antiproton trajectories any of the target nucleons will 
change many times its momentum. 
In presence of Coulomb attraction, 
the previous eq.(\ref{free11}) is wrong, and 
must be rewritten as:  

\arr
d\sigma_{A,N}(\vec p,0,\vec \pi_j)
\ =\ \int {{d^3k}\over {(2\pi)^3}} S(k,k) 
{{F_Z(\vec p,\vec k)\beta(\vec p,\vec k)} \over 
{F_1(\vec p,0)\beta(\vec p,0)}}
d\sigma_{1,N}(\vec p,\vec k,\vec \pi_j),
\label{free12}
\endarr
where the function $F(\vec p,\vec k)$ depends on $\vec p$ 
and $\vec k$ via the relative velocity 
$\beta(\vec p,\vec k)$, 
and satisfies the two limits $F(\beta)$ $\rightarrow$ 
$Z/\beta$ for small 
$\beta$ (practically for $\beta$ $<$ 0.05)
and $F(\beta)$ $\rightarrow$ $1$ at large $\beta$ (we know already 
that, for momenta of some hundreds MeV/c, $\sigma$ $\sim$ $1/\beta$
as suggested by eq.(\ref{free11}) and by the general theory 
of inelastic processes in absence of Coulomb effects).

Some more general observations are necessary: 
First, our ``uncharged antiproton'' cross sections 
can't be assumed to be exactly equal to antineutron 
cross sections on a proton target, because the $\bar{p}p$ 
and $\bar{n}p$ states have a different isospin composition:
$\bar{n}p$ is a pure $I$ $=$ 1 state, while $\bar{p}p$ mixes 
$I$ $=$ 1 and $I$ $=$ 0 states. Working the other way round, 
by comparison 
with data one can extract the separate isospin contributions. 
On the other side, if the target is an isoscalar nucleus 
then our ``uncharged antiproton'' is equivalent to 
an antineutron. 

Another relevant point 
is that in the case of a free neutron target 
there should be no Coulomb focusing effect. 
However, since experiments 
on $n\bar{p}$ scattering are normally 
performed on deuteron targets, the $n\bar{p}$ cross section 
is as much ``Coulomb focused'' as the $p\bar{p}$ one. 
So, in the following, if ``free neutron target 
cross sections'' are more 
or less explicitely used, they must be meant as extracted 
from $\bar{p}d$ scattering. Obviously, the cross section 
for $\bar{n}p$ scattering (which is an actually performed 
experiment\cite{arm,feli}, and 
where strong interaction are the same as 
in $\bar{p}n$ scattering) will be another thing, 
directly connected to the cross section for 
$\bar{p}$ scattering on a really free neutron. 
In practics, however,  
we will not distinguish between proton and neutron 
interactions in this work. 

In principle, Coulomb interactions affect the final state 
also. But, since such effects are of practical importance 
when the velocity of the involved particles is $\beta$ $<<$ 
1, final state Coulomb effects would be relevant only when 
the total mass of the final state is $\approx$ 2$M_p$, which 
seems to be a very rare event\cite{af1}. 

\section{Some applications}

First, some relevant limits have to be considered. For 
incoming $\bar{p}$ momenta sensibly larger than the 
Fermi momentum, i.e. sensibly larger than 200 MeV/c, 
$\beta(\vec p,\vec k)$ $\approx$ $\beta(\vec p,0)$ 
and $d\sigma_{1,N}(\vec p,\vec k,\vec \pi_j)$ $\approx$
$d\sigma_{1,N}(\vec p,0,\vec \pi_j)$, so that exploiting 
the normalization rule of $S(k,k)$ we simply get 

\arr
\sigma_{A,N}(p)\ \approx\ 
A \sigma_{1,N}(p),\ p\ >>\ 200\ MeV/c
\label{free13}
\endarr
for the total annihilation 
cross sections on nuclear and proton targets. 
The limit of the PWIA treatment 
is evident: no eclipse effect (related with multiple 
scattering and leading to a more realistic $A^{2/3}$ 
dependence\cite{aldo}) is present in this result. 

Another interesting limit is at the opposite of 
the scale, for incoming $\bar{p}$ momenta clearly 
below the Fermi momentum, so for $p$ $<<$ 100 MeV/c. 
Then eq.(\ref{free12}) can be approximated by 
\arr
d\sigma_{A,N}(\vec p,0,\vec \pi_j)
\ =\ {Z \over \beta^2(\vec p,0)} 
\int {{d^3k}\over {(2\pi)^3}} S(k,k) 
\beta^2(0,\vec k)
d\sigma_{1,N}(0,\vec k,\vec \pi_j). 
\label{free14}
\endarr

Now one can exploit the fact that 
$\beta^2(0,\vec k)
d\sigma_{1,N}(0,\vec k,\vec \pi_j)$ is a regular function,
dominated by the S-wave contribution for not too large $k$. 
Then we may write

\arrn
\beta^2(0,\vec k)
d\sigma_{1,N}(0,\vec k,\vec \pi_j)\ =\ 
C \vert T(0,k,\pi_j)\vert^2 
(2\pi)^3\delta^3(\vec k -\sum\vec \pi)
\delta(E_{fin}-E_{in})
\prod {{d^3\pi_j}\over{(2\pi)^3 2E_j}}\  
\approx
\endarrn
\arr
\approx \ 
C \vert T(0,0,\pi_j)\vert^2 (2\pi)^3\delta^3(\vec k -\sum\vec \pi)
\delta(E_{fin}-E_{in})
\prod {{d^3\pi_j}\over{(2\pi)^3 2E_j}} 
\label{free15}
\endarr
and

\arrn
d\sigma_{A,N}(\vec p,0,\vec \pi_j)\ =
\endarrn
\arr
=\ {{CZ} \over \beta^2(\vec p,0)} 
\int {{d^3k}\over {(2\pi)^3}} S(k,k) 
\vert T(0,0,\pi_j)\vert^2 
(2\pi)^3\delta^3(\vec k -\sum\vec \pi)
\delta(E_{fin}-E_{in})
\prod {{d^3\pi_j}\over{(2\pi)^3 2E_j}}
\label{free16}
\endarr

Substituting $\vert T(0,0,\vec \pi_j)\vert^2$ with some average on 
relative pion directions $\vert T\vert^2$ 
(with the final aim of extimating 
total cross sections),  and integrating on the pion 
phase space,  
the previous two equations become 

\arr
\sigma_{1,N}\ 
\approx\  
{1 \over {\beta^2(\vec p,0)}}
C\vert T\vert^2 \int 
(2\pi)^3\delta^3(\sum\vec \pi)
\delta(\sum E_j-2M_p)
\prod {{d^3\pi_j}\over{(2\pi)^3 2E_j}} 
\label{free15b}
\endarr

\arr
\sigma_{A,N}
\ \approx\ {Z \over \beta^2(\vec p,0)} 
C \vert T\vert^2 
\int S(k,k)_{\vec k = \sum\vec \pi}
\delta(\sum E_j-2M_p)
\prod {{d^3\pi_j}\over{(2\pi)^3 2E_j}}
\label{free16b}
\endarr
(approximately valid for $p$ $<<$ 100 MeV/c only).

The two reasons that justify extraction of an averaged 
$\vert T\vert^2$ are: 1)  now we are interested in total 
cross sections, so fine correlation effects between 
the momenta of the emitted pions don't concern us too much,  
2) we want to compare total cross sections with different targets 
and we expect that in the ratio the effects related with 
$\vert T\vert^2$ are largely cancelled.  
Of course this procedure would hide the effect of a 
possible resonance peak of the $\bar{p}p$ cross section 
decaying into a specific channel. 
So we roughly obtain 

\arr
{{\sigma_{A,N}}\over {\sigma_{1,N}}}\ 
\approx\ 
Z {{\int [S_A(k,k)]_{\vec k = \sum\vec\pi} d\Phi_N} \over
{\int (2\pi)^3 
[\delta^3(\vec k)]_{\vec k = \sum\vec\pi} d\Phi_N}}
\ \equiv\ Z R_A,
\ \ p\ <<\ p_{Fermi}
\label{free17}
\endarr

The fact that this equation only contains nuclear quantities 
clearly shows its advantages and limitations. We can 
say that $R_A$ 
gives some general information on the way the nuclear 
structure may affect a largely esothermic 
reaction between a low energy hadronic 
projectile and a nucleon, with large 
(assumed) independence from the specific character of this 
reaction. 
The calculation of the integral in the numerator 
of $R_A$ is not easy in general, even 
numerically. In the peculiar case of two pions only in the final 
state, four of the six integrations can be performed analitically
(there is full spherical symmetry in the emission of one of the 
two pions, at least in the limit $p$ $\rightarrow$ 0, 
and azimuthal symmetry for emission of the second pion;
the integration over the energy of one of the two pions is removed 
by the energy shell condition). For the peculiar choice 
of a Gaussian distribution 

\arr
S(k,k)\ \equiv\ A (2\pi)^3 {r^3 \over {\pi^{3/2}}}  e^{-r^2k^2}
\label{gauss1}
\endarr
the integral in the numerator of $R_A$ 
can be fully calculated with the approximation 
of the steepest descent. 

\begin{figure}[htp]
\begin{center}
\mbox{
\epsfig{file=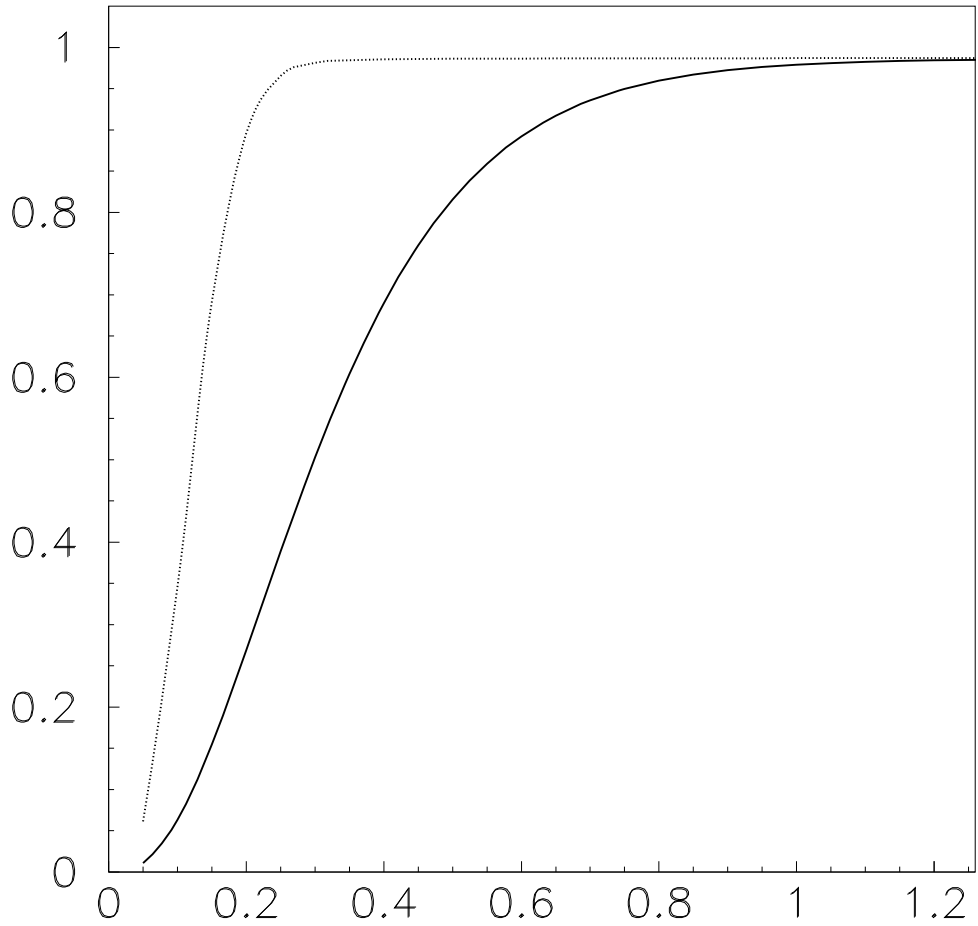,width=0.9\linewidth}}
\end{center}
\caption[]
{\small\it \label{fig4}
Continuous line: the ratio $R_A/A$ (see text) calculated
in the two simple cases of a Gaussian distribution (dotted curve)
and exponential distribution (continuous curve). $R_A/A$ is 
presented as function of the distribution parameter $r$ (fm). 
}
\end{figure}

\noindent
Direct numerical integration confirms the 
validity of this approximation for $r$ $>$ 0.4 fm: for 
such values of $r$ the integral is dominated by a narrow 
kinematical region around 
$\vert \vec \pi_1\vert$ $=$ $\vert \vec \pi_2\vert$, leading 
to a value of the integral which is $r-$independent. 
The result is that $R_A$ $\simeq$ $A$ 
for any reasonable value of the 
parameter $r$ (for any $r$ $>$ 0.4 fm). For 
unrealistically small $r$ the ratio tends to zero.  
For large enough nuclei ($^{12}$C onwards)
$r$ simply means $1/p_{Fermi}$ $\sim$ 1 fm. For deuteron 
$r$ would be larger, although in this case a simple exponential 
shape for $S(k,k)$ 
is more reasonable. When the deuteron single 
particle momentum distribution $N(k)$ 
is represented in a semilog plot\cite{bjnz}, it is 
almost a straight line with negative derivative $\sim$ 
$-6$ between $k$ $=$ 0 and the $S-D$ interference minimum.
If we choose the simple exponential law 

\arr
S(k,k)\ \equiv\ A {r^3 \over {8\pi}}  e^{-rk}
\label{exp1}
\endarr
instead of the previous Gaussian one, 
a numerical calculation again gives 
$R_A$ $\simeq$ $A$ for $r$ $>$ 1 fm 
(3 percent of difference at $r$ $=$ 
1 fm). At smaller $r$ the ratio $R_A$ is smaller than 
$A$, and the convergence to the limit $A$ 
is slower than in the Gaussian case: $R_A$ $\simeq$ 
0.7 for $r$ $=$ 0.4 fm and 0.8 for $r$ $=$ 0.5 fm. 
But, for any 
reasonable value of $r$ we get the most obvious result. 
For deuteron $r$ $\approx$ 6 coincides approximately with 
twice the radius of the deuteron $S-$wave wavefunction
(in deuteron $S(k,k)$ $\equiv$ $\vert\psi(k)\vert^2$). 
These results, showed in fig.4,  
descourage us from trying to implement a 
Montecarlo integral to estimate $R_A$ with a larger number 
of emitted pions. The feeling one gets from the above 
calculations is that the PWIA calculation does not reserve special 
surprises, at least for those values of $r$ 
which are reasonable in mean-field models. 
For a discussion of 
the role of small$-r$ components in $S(k,k)$ (coming
from non-mean-field contributions) 
see the next section.  

A ``surprise'' could arrive from the existence of 
a clear resonance peak 
in the $\bar{p}p$ cross section in the energy 
region which is just below the lower limit of the 
explored range. In the nuclear case 
Fermi motion would spread the effect of such a peak through a 
wide kinematical range, with relative 
advantage of the $\bar{p}p$ 
cross section in the energy region which is close to the 
resonance peak, and of the $\bar{p}-$nucleus 
cross section at larger energies. This effect can 
be estimated by  eq.(\ref{free11}) or 
(\ref{free12}), it is well known in many different 
branches of nuclear physics and it is just a form 
of Doppler effect. In the following we show an 
example of such Fermi motion spread of a resonance 
peak. On the other side, we notice that it is 
possible to fit the presently 
available data on $\bar{p}p$ annihilation with 
curves that produce 
a progressive trend towards the above quoted 
non resonant $1/p^2$ law, 
without need of additional contributions. 
Written with respect to the momenta $\vec p$ and 
$\vec k$ of the colliding $\bar p$ and $p$ 
a resonance Breit-Wigner peak becomes (in nonrelativistic 
approximation and assuming onshellness of both particles) 

\arr
\beta^2 d\sigma\ =\ {C \over 
{\big(\vert \vec p - \vec k\vert - p_o\big)^2 + B^2}}
f(\vec p, \vec k, \vec \pi) d\Phi_N,
\label{res1}
\endarr
where $B$ corresponds to $\Gamma/2$ in the momentum space, 
$C$ depends slowly on kinematics,
$p_o$ is the momentum corresponding to the resonance energy 
for target at rest, 
and $f$ depends on the specific angular momentum of the 
channel where the resonance is present.
The Breit-Wigner denominator contains the total center of mass 
energy of $\bar{p}$ and $p$. Since one of the two 
colliding particles is 
bound to a nucleus this concept is ill-defined, and some 
prescription for off-shellness treatment is necessary. We 
have adopted the easiest form, i.e. to treat the proton as a free 
particle with nonrelativistic energy $m+k^2/2m$. Another 
possible prescription, which expresses the exactly opposite 
point of view, is to fix the proton energy to its mass 
plus a small negative constant, for any  
$\vec k$. At a qualitative level, results don't depend 
too much on the choice of the offshellness prescription,
since the center of mass energy is anyway strongly dependent 
on the product $\vec p\cdot \vec k$, which is present 
in both cases. 

Neglecting the differences coming from the phase space 
integration of the channel function $f$, 
from eq.(\ref{free13}) we see that the 
total annihilation cross sections for $\bar{p}p$ and 
$\bar{p}A$ will be proportional to the two 
comparable factors: 

\arr
\beta^2 \sigma_1\ =\ {C \over 
{(\vec p - p_o)^2 + B^2}}
\label{res2}
\endarr

\arr
\beta^2 \sigma_A\ =\ Z \int {{d^3k}\over{(2\pi)^3}}
{C \over 
{\big(\vert \vec p - \vec k\vert - p_o\big)^2 + B^2}}
\label{res3}
\endarr
(we work assuming full low energy effect of the Coulomb 
interaction). 
As an example, in fig.5 we show the two curves 
(\ref{res2}) and (\ref{res3}) calculated 
numerically with 
gaussian $S(k,k)$ $=$ $A(r/\sqrt{\pi})^3exp(-r^2k^2)$, 
with a choice of resonance parameters 
$B$ $=$ 10 MeV/c, $p_o$ $=$ 20 MeV/c, and  
nuclear parameters $r$ $=$ 2 fm, $A$ $=$ 4 and $Z$ $=$ 2. 

\begin{figure}[htp]
\begin{center}
\mbox{
\epsfig{file=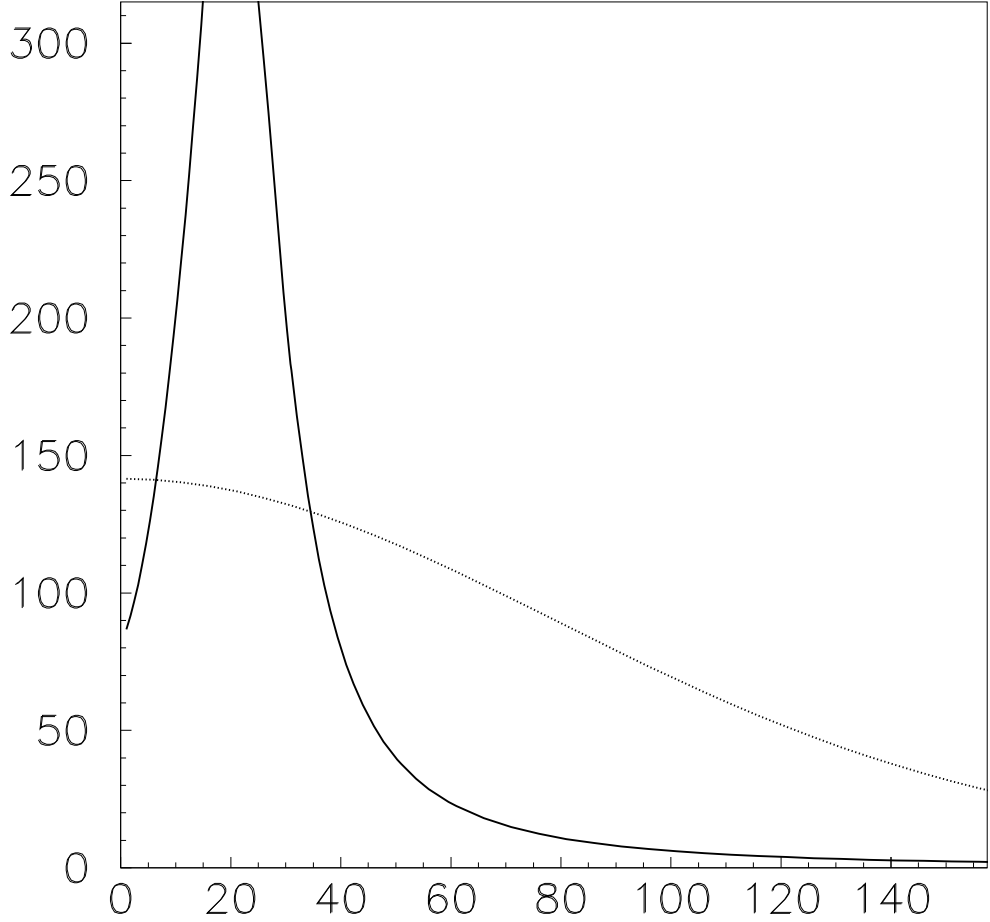,width=0.9\linewidth}}
\end{center}
\caption[]
{\small\it \label{fig5}
Nuclear smearing of a resonant contribution in the cross section
for $\bar{p}p$ annihilation into a given channel. The 
continuous and dotted curve represent eq.(\ref{res2}) 
and eq.(\ref{res3}) respectively (in mb), versus p (MeV/c). 
We have chosen $B$ $=$ 10 MeV/c,
$p_o$ $=$ 20 MeV/c, represented the nucleus via a Gaussian 
distribution with parameters corresponding to $^4$He (see text
for details). The vertical scale 
is the same for both curves and corresponds to the 
choice 
$C$ $=$ $4\cdot 10^4$, for $\sigma$ expressed in mb and 
$p_o$ and $B$ in MeV/c (equivalent to $C$ $=$ 1 for $p_o$ 
and $B$ expressed in 1/fm).}
\end{figure}

It is easy, although long, to demonstrate approximately 
the obvious fact that at small $\vec p$ and for 
$1/B$ $>>$ $r$ (that means 
$B$ $<<$ $p_{Fermi}$) the peak of the resonance is 
enlarged to a size $\sim$ $p_{Fermi}$. One can approximate 
the Breit-Wigner shape $1/[(p-p_o)^2-B^2]$ 
with a Gaussian curve 
$exp[-(p-p_o)^2/B^2]/B$. This approximation is good 
near the resonance peak, and as wrong as a Breit-Wigner 
far from the peak (where, however, the fall of the 
nuclear momentum distribution 
kills contributions in both cases). Then substituting this 
approximation in eq.(\ref{res3}), together with neglecting 
terms of second order with respect to the small parameters 
$B/r$ and $p/k$ allows for calculation of the integral,
which is roughly equal to a constant times $exp(-r^2p^2)$. 

Evidently in the figure we see something that can  
be exchanged for an inversion behavior if the data 
stop at 30 MeV/c. However, this such a peculiar situation  
is unlikely. In fact, 
we tested three further possibilities: (1) a larger 
resonance peak; 
(2) a set of many partially overlapping resonances.  
(3) a set of many clearly spaced resonance 
peaks. In case (1), for $B$ larger than 
a few tenth MeV/c the nuclear target cross section overcomes 
everywhere the single nucleon target one. So a phenomenon 
like the one shown in fig.5 is not seen but for 
really narrow resonances. In case (2) the behavior reproduces 
what one could expect from the analysis of the single resonance 
case: If the complete set of resonance peaks 
in $\bar{p}-$nucleon annihilation clearly rises 
from the background and is contained inside a narrow range 
(within a few tenth MeV/c) then the behavior is like in fig.5. 
If the set of resonance peaks of the $\bar{p}-$nucleon 
annihilation spreads its effects out of this range, 
then the situation is like in example (2), i.e. like the 
case of a single resonance with $B$ $>>$ 10 MeV/c. In the limiting 
case where the range occupied by the overlapping resonances is of 
the size of the Fermi momentum or larger 
we simply find the same ratio 
between $\bar{p}p$ and $\bar{p}A$ 
cross sections as in the nonresonant case. 
In case (3) the nuclear Fermi motion completely 
cancels any chance to distinguish different resonances.
Even in the simple case of two narrow peaks only, if their 
distance is less than 200 MeV/c it is impossible, 
for one who looks at $\bar{p}-$nucleus data, to imagine 
that two resonances are present in the $\bar{p}-$nucleon 
cross section. A possible 
limiting case could be the one where in a wide region 
(of size larger than 100 MeV/c)  
many resonances are present in the $\bar{p}-$nucleon 
cross section, and although they are clearly spaced 
it is not experimentally 
possible to resolve them: then as in case (2) one 
will see two continuous curves whose ratio $R_A$ 
is the same as predicted for the nonresonant 
background. As a conclusion, an inversion behavior 
as the one experimentally seen 
at the lowest momenta of the explored range could be justified 
in terms of one or a few resonance peaks, but they all 
should be narrow ($B$ not much larger than 10 MeV/c) 
and their peaks 
should be all confined in the region $p$ $<$ 30 MeV/c 
and very clearly emerge from the background. 

One more point should be stressed: in the averaging 
procedures leading to eq.(\ref{free17}) the effect of 
any irregularity in the $\bar{p}-$nucleon $\vert T\vert^2$ 
is lost. Such an equation could not be used to show 
how a resonance peak is spread by Fermi motion. 
However, the example of 
the resonance peak suggests that any variation 
in $\vert T \vert^2$ can 
be the cause of an inversion behavior (originated by 
Fermi motion spreading) if the relative change of 
$\vert T\vert^2$ is large within a momentum range 
of a few tenths MeV/c. 

\section{Over Inpulse Approximation and Mean Field: a 
discussion} 

The fact that PWIA, within mean field approximations, 
does not 
explain inversion means that what we see is not a 
situation where annihilation takes place on one target 
nucleon only, with the remaining $A-1$ just acting 
as spectators. Annihilation probably takes place on 
one nucleon mainly, however the other ones do play 
some more or less direct role in depressing its 
effectiveness. Here 
we limit ourselves to explainations involving nuclear 
physics in traditional sense, i.e. without considering 
phenomena related with the quark structure of a 
nucleon cluster. The common point is that both the 
consideration of components in S(k,k) related with 
momenta $k$ $>>$ $p_{Fermi}$, and the multiple scattering, 
are processes which directly 
involve more than one target nucleon, and 
have a large probability of leaving the 
residual nucleus in an excited or fragmented state. 
The data\cite{obe1,obe2} on the ``inversion'' region 
don't classify final residual states. In addition, they 
involve Deuteron or $^4$He only, so a systematics 
of the behavior of the residual is not possible yet. 
A knowledge of the distribution of the residual 
states would be decisive in this respect. 
In the case of non IA processes (i.e. involving multiple 
interactions) 
a second scattering process accompanying the  
annihilation would transfer a large recoil momentum 
to a single spectator nucleon, increasing the probability  
of breaking the residual. In the case of an annihilation 
process within IA (i.e. directly 
involving a single target nucleon) but 
picking up a nucleon with initial momentum 
$k$ $>>$ $p_{Fermi}$, 
this large 
momentum of a single nucleon is probably due to direct 
interaction with another nearby nucleon. This 
spectator nucleon would be left with large recoil momentum, 
just as in the rescattering case. 

Let us examine first the role of S(k,k): 
The Gaussian and exponential simple models we have used above 
are clearly approximations of distributions of 
mean-field character. 
However, the slower convergency of the results we get 
when a simple exponential law is used instead of a gaussian one 
suggests a certain dependence on the high momentum part of 
the distribution $S(k,k)$, i.e. for $k$ $>$ $P_{Fermi}$. 
At $k$ $\sim$ 1/(0.5 fm), i.e. $k$ $\sim$ 
400 MeV/c, $S(k,k)$ receives large contributions by 
nuclear surface effects and short range correlations. 
The first effect can enter a mean field shell treatment, 
but not when gaussian functions are used, since no surface
thickness parameter is present. The second 
one requires sophisticated nuclear models, beyond 
the mean field approximation\cite{bgp}. 
In practics the relevant points are that such 
effects depend on $A$ in a different way, 
and carry different dimensional scales.  
The mean field function introduces into any Fourier 
transform a large component at $k$ $\sim$ 
$R_{nucleus}^{-1}$ 
$=$ $1/(r_oA^{1/3})$, $r_o$ $\simeq$ 1.3 fm. 
So one may imagine a kind 
of $A e^{-br_oA^{1/3}k}$ ($b$ $\sim$ 1) dominating term. 
Surface effects 
and direct short range nucleon$-$nucleon interactions 
produce large components which are relevant at $k$ 
$\sim$ $1/r_c$ or $1/r_s$, where the short range 
parameters $r_c$ and $r_s$ are smaller than 1 fm 
(typically: 0.5 fm)\cite{bnr}. 
The surface term could be $\sim$ $A^{2/3}e^{-br_sk}$ 
($b$ $\sim$ 1), while the repulsion term 
is $\sim$ $A^2e^{-br_ck}$ (i.e. proportional to the number of 
couples). 
Clearly, it is like having, in a nucleus, different 
momentum distributions 
living aside, and sharing a total 
normalization factor 1. In the case of deuteron 
the high momentum part of the momentum distribution 
coincides with the D-wave. In more complex nuclei, 
the calculation of the spectroscopic factors 
in $(e,e'p)$ reactions\cite{bgp} 
suggests that not much more than 60\% of $\int S d^3k$ is 
filled by the mean field contributions. So a large part 
of $S(k,k)$ could be represented by terms damping on 
a scale $r$ $\sim$ 0.5 fm, where the above ratio $R_A$ 
has been estimated to be smaller than $A$. However, 
fig.4 shows that $R_A$ $\simeq$ 0.8 for $r$
$=$ 0.5 fm, so that 
even taking into account a 40\% contribution by a 
term like 
$S$ $\approx$ $exp(-r k)$ with $r$ $\simeq$ 0.5 fm, 
the overall ``large momentum decrease'' 
would be contained within 10 \%. 
To estimate with more precision  how much could be 
the real weight of such large $k$ terms in the phase 
space integral for the total annihilation cross section, 
a first interesting step could be a model calculation 
of the relative amounts of annihilations involving the 
deuteron S and D components. 
Experimentally it does not seems easy 
to decide whether residual fragmentation  
(whenever found) originates 
in initial state or in multiple scattering effects.   
 
When multiple scattering effects are 
introduced the IA results could be largely
modified. A long series of possible mechanisms should 
be studied, and we are doubtful about the possibility 
to have all of them under control within short time 
since now. Within the limits of our fantasy and 
knowledge, we list some of the possibilities: 

1) Elastic shadowing, i.e. 
the eclipse effect between the nucleons.  
This effect is surely present although 
its role at low energies is not so intuitive as in the 
corresponding high$-$energy problem\cite{glau}, 
where it may be identified with the decrease of 
the reaction probability on a target nucleon 
due to flux removal in the previous reactions with 
another one. 
At low energy the $\bar{p}$ position is not well defined 
inside the target nucleus, due to $\Delta x$ 
$\sim$ $1/\Delta p$ $\sim$ $1/p$ $\rightarrow$ $\infty$. 
So we can't visualize easily the shape of the ``shadow'' 
that the first target nucleon casts on the second one. 
We discuss below the possibility of treating 
this mechanism within DWIA. 

2) Inelastic shadowing\cite{gri}, i.e. 
the intermediate (real or virtual) 
mesonic state created by an initial 
annihilation is converted into $\bar{p} + p$ 
by the opposite process in a second step.  
This transfers a part of the inelastic rate into the 
elastic one. The only sensible way to treat this 
mechanism could be via a coupled channel 
model\cite{pro1}. 

3) Inelastic antishadowing: in the first 
inelastic event a state is created which is not allowed 
by the $overall$ conservation rules, although it may 
temporaneously exist as a virtual state. 
A second scattering 
event converts this virtual state 
into a real and allowed one, completing the 
annihilation process. 
One can speak of multistep annihilation 
processes, peculiar of nuclear targets. It does not seem 
easy to make good quantitative predictions on its 
effectiveness. Again, a coupled channel model could be 
the method. 

4) Soft  distortions of the projectile wave, produced 
by coherent interactions with the nuclear matter. With 
``soft'' we mean that they can involve 
excitation of the spectator set of $A-1$ nucleons, but 
not drastic changes of the internal state  
of any of the individual nucleons. These 
interactions could perhaps be treated via DWIA with 
traditional nuclear optical 
potentials with a small imaginary part. In the 
momentum space, 
they transform the $\bar{p}$ wavefunction $\psi(k)$ 
from a plane wave $\delta(k-p)$ into a distribution 
with peak at $k$ $\approx$ $p$ and fluctuation $\sim$ 
200 MeV/c. Intuition suggests that their effect should 
consist mainly in enhancing the kinematical spreading  
which is already intrinsic in PWIA. 

5) Soft distortions of the projectile wave, produced 
by diffraction accompanying 
hard inelastic $\bar{p}-$nucleon events. In practics, any 
annihilation creates a ``vacuum region'' in the projectile 
wavefunction $\psi$, and the borders of this region are 
characterized by large space derivatives of $\psi$, 
meaning high momentum components of the initial state 
wavefunction. About the possibility to treat this 
by DWIA, see below. We must remark that, as the analysis 
of the black sphere model contained in the next section 
shows, within that model this is the main 
cause of depression of the annihilation rate. 

6) Soft distortions of the projectile wave $\psi$ 
produced by the above mechanism (2). Let us imagine 
that a $\bar{p}+p$ couple 
is converted by the first annihilation process into 
a set of 2 pions in a position $\vec r_1$, and by 
the opposite process into a couple $\bar{p}+p$ 
again, in the position $\vec r_2$. 
In nuclear matter the propagation 
properties of the two pion system from $\vec r_1$ to 
$\vec r_2$ are different from the corresponding 
ones of the $\bar{p}+p$ system. In the high energy 
regime it is relatively easy to calculate the 
phase difference associated with this mechanism\cite{gri}
once one knows the mass of the intermediate state. 
We don't know a corresponding low energy theory. 

7) Soft final state interactions (SFSI)
of the meson system 
(the hard ones can be included into some of the previous 
cases). SFSI can be treated approximating 
the final meson plane waves with damped waves in Glauber 
style. These express the fact that any precisely identified 
meson state will be partly depleted by interactions 
with the residual nucleus. Although probably SFSI do not 
change much the total annihilation rate, it 
will enhance the number of annihilations accompanied 
by fragmentation of the residual. It will also 
introduce high 
momentum components into the meson wavefunctions. Due 
to the fact that the meson waves are already high momentum 
ones, this should be less relevant than the corresponding 
initial state mechanisms (4) and (5). 

8) Multiple interactions between $\bar{p}$ and 
the same target nucleon. If these interactions are 
consecutive they are reabsorbed in the 
matrix element $T$ for the $\bar{p}-p$ interaction 
used in the IA. If they are not consecutive they 
escape IA. 
At very low energy, because of delocalization of the wave 
functions, multiple interaction events on the same nucleon 
and backward scattering can be relevant. E.g. triple 
scattering on deuteron is not necessarily a rare event. 

A common element which is present in 
some of the previous cases 
is the distortion of the incoming or outgoing waves, 
due to hard or soft interactions with the nuclear matter. 
In the target region this creates Fourier components of 
the wavefunction of 
$\bar{p}$ and of the final mesons which are not 
present asintotically.
So, even assuming a given vector $\vec k$ for the nuclear  
proton, the kinematical details of the  
annihilation are unknown. E.g. in the predictable 
case of strong absorption on the nuclear 
surface, the large gradients which accompany 
the fast damping of the $\bar{p}$ wavefunction 
will introduce high momentum components into it, 
partly shifting 
the reaction to a region where it can't be considered as 
a ``low energy'' process. 

In large nuclei some of the 
above described effects could be 
treated by DWIA. In this formalism 
the waves participating 
to the annihilation are not plane waves, but damped 
or somehow distorted 
waves. So, when interacting with one given nucleon, the 
projectile wave takes into account the flux removal, or 
the wave distortion, due to 
possibility that the $\bar{p}$ (or the final state mesons) 
have already interacted (or will interact) with  
another nucleon. Leaving aside formal problems, 
this treatment can be justified 
when it is sensible to write the interaction hamiltonian as 
a sum of two terms $H_1$ $+$ $H_2$, where $H_1$ produces the 
matrix element $T$ used in the IA, and $H_2$ is 
responsible for the wave distortion. Supposedly, in our 
case $H_1$ would be the interaction between $\bar{p}$ 
and one nucleon, and $H_2$ the interaction between 
$\bar{p}$ (or the final mesons) 
and all the other ones. In practics and with all 
kinds of hadronic projectiles, DWIA 
normally means either the use of the Glauber 
formalism, or a distortion of the wavefunction by a 
suitably fitted optical potential. It can  
work as far as the average nuclear matter interactions 
with the projectile are known, which is absolutely not our 
case (they are exactly what we would like 
to understand here). In addition this kind of 
implementation of DWIA 
would mean an unpleasant mixing between two models, 
the IA and the optical potential model, in a regime 
where still both of them 
need separate analysis and understanding. 

A proper study of inelastic shadowing 
or antishadowing is difficult, even at qualitative level, 
and would require a multi-channel 
model. In the simplest case\cite{pro1} one only takes into 
account two channels (elastic and ``effective inelastic''). 
Although this model can  
produce strong inelastic shadowing and inversion\cite{pro1}
the quantitative results depend critically 
on the coefficients (and phases) 
one chooses for the transition 
amplitudes between the elastic and the excited channels. 
Physical constraints over these coefficients are  
presently poor, making a quantitative analysis of 
low energy inelastic shadowing a difficult task. 
As far as only total annihilation rates are known, 
a many channel model is equivalent to the 
optical potential model. In the latter case 
one sums effectively over all the 
transitions (through both elastic and inelastic 
channels) leading to the elastic channel in the final 
state. All the information about the details of the 
inelastic channels is hidden in the optical potential 
(whose form can be 
formally derived starting from a set of coupled channel 
equations). 
In the many channel model one has to introduce specifical 
information about the inelastic channels. If 
this information is arbitrary the two models 
are in practics equivalent, although 
one can gain some 
physical interpretation of the phenomena hidden behind 
the optical potential. 

Probably the easiest starting ground for a study 
of multiple interactions is a deuteron target. 
As we have anticipated, support to the role of 
non IA terms in the low energy interactions between 
$\bar{p}$ and nuclei 
comes from an analysis of the problem 
of the width and shift of the 
levels of the antiprotonic deuterium atom\cite{wic}. 
In that work consistent cancellation has been found  
between effects which can be associated to single and double 
scattering. 
To complete the picture, recently we have 
been informed of a nonperturbative 
three-body analysis of $\bar{p}D$ 
scattering\cite{pro2}. 
In this work some reasonable forms for the 
$\bar{p}-$nucleon interaction have been postulated 
and 3-body equations solved for the system $p+n+\bar{p}$. 
It has been found that an 
annihilation $\bar{p}D$ 
cross section smaller than the $\bar{p}p$ 
one cannot be obtained with a purely absorbitive 
$\bar{p}-$nucleon interaction. 
It seems that more complex mechanisms are needed. 
This would decrease both the role of multiple 
interaction effects and of the optical model in its 
simplest form. 

The obvious conclusion of this section is that, 
due to the difficulty in taking into account all the 
previously listed effects, these aspects of the problem 
will probably remain open for a long time. 

\section{The black sphere optical potential model}

In ref.\cite{pro1} a black sphere potential 
$V(r)$ $=$ $-iW$ for $r$ $<$ $R$ 
was used to show that, increasing $W$, one finds  
an initial increase followed by a definitive 
decrease of the annihilation rate, which tends to zero 
when $W$ $\rightarrow$ $\infty$  
(at very low energies). We have carried on 
some further calculations on the 
same easy model, which   
showed us that if one increases $R$ at constant $W$, 
the annihilation rate saturates fastly, leading to 
a constant S-wave contribution to the cross section. 

Trusting this potential model would imply to expect 
the same behavior when comparing $\bar{n}-$nucleus 
cross sections at different mass numbers.
We could associate each nucleus with an imaginary  
potential well with radius  
$R$ $=$ $A^{1/3}r_o$, $r_o$ $\approx$ 1.3 fm, and assume 
$W$ $=$ (unknown) constant for all nuclei.  
The last assumption means 
to consider the strength $W$ as a local mean property  
of nuclear matter. Nuclear matter local properties 
are supposed to be similar in most of the nuclei (with the 
exception of the lightest ones). 
Then the consequence of the above reported 
saturation effect 
would be an $A-$independent $\bar{n}$ annihilation 
cross section at low energies,  
and a $\sim$ $Z$ $\sim$ $A/2$ dependence for $\bar{p}$
annihilation (due to the Coulomb corrections). 

To understand how much of this potential model can be 
general, we have analyzed  
the behavior of the wavefunction $inside$ the black sphere. 
As we explain a little more in detail below, 
we find that exponential 
damping of the $\bar{p}$ density 
inside the black sphere 
happens on a scale $R-r$ $\sim$ $1/\sqrt{MW}$ 
(in natural units; $M$ is the 
projectile reduced mass $\sim$ 1 GeV, $W$ is the strength 
of the imaginary part of the potential; we assumed zero 
real part). So, for any 
potential radius $R$ sensibly larger than $1/\sqrt{MW}$ 
all of the incoming isotropical 
flux disappears before reaching the 
center of the target. This introduces a saturation 
mechanism, of the same kind that at higher energies causes 
the transition from volume absorption to surface absorption. 

For better understanding the peculiarities of the low 
energy problem, we would like to start by 
analyzing some 
mathematical properties that, in the $high$ energy problem,
lead to the ``obvious'' association between strong 
flux absorption 
by the target and a high reaction probability. 
Some of these key properties are not present in the 
corresponding problem at low energies. 

First, the ideal geometry of the problem is 
spherical at low energies, and axial 
(with defined orientation) at large 
energies. From a certain point of view, 
the lowest possible border of the ``high energy'' region,
(where a flux direction with back-front orientation can 
be clearly recognized inside the nuclear volume) 
may be associated with the presence of a consistent $P-$wave 
contribution. In the $S-$wave dominance regime, 
there is no preferred direction, and concerning orientation 
we can only speak of ``inward'' and ``outward''. But with 
both S-wave and P-wave present with similar relevance, the 
direction of the nonzero average angular momentum 
identifies a preferential $\hat z-$axis 
in space, and the interference between the two waves 
(S $\sim$ 1, P $\sim$ $cos(\theta)$) creates a difference 
between the $\theta=0$ and $\theta=180^o$ values of 
$\vert \psi\vert^2$, which allows for identifying an 
orientation of the $\hat z-$axis. 
At really large energies, 
where many partial waves enter the scattering region
($l_{max}$ $\sim$ $Rp$ in natural units)
the full projectile wavefunction $\psi$ tends to the form  
$e^{ipz}f(\vec r)$, where $\hat z$ is the 
incoming flux direction and $f(\vec r)$ depends on $\vec r$ 
slowly enough to allow one to consider $f$ as a constant 
over a space region of size $1/p$. 
Assuming the nuclear center to be in the origin, 
with such a form we can 
say that $\psi$ enters the nucleus on one 
side, and exit from the other side. 
Roughly, we can speak of an ``entrance'' 
surface portion ($z$ $<$ 0)
and of an ``exit'' surface portion ($z$ $>$ 0). 
On both sides the values of the $external$ wavefunction 
must match the values of the $internal$ one. 
The internal function assumes rather different 
values in the two regions: 
large, at the ``entrance'', small, at the ``exit''. 
This can be seen exploiting the eikonal approximations:  

\arr
\nabla^2 + k^2\ \approx\ 
{{\partial^2} \over {\partial z^2}} + k^2 \ =\ 
\Big({{\partial} \over {\partial z}} + ik\Big) \ 
\Big({{\partial} \over {\partial z}} -ik\Big) \ 
\approx\ 2ik\Big({{\partial} \over {\partial z}} -ik\Big). 
\label{eiko1}
\endarr 
These approximations reduce the Schroedinger equation 
to a first order simple 
equation with solution $exp(-mWz/k)$
in a region characterized by the uniform potential 
$V$ $=$ $iW$ (the other possible solution 
$exp(+mWz/k)$ is discarded 
on the basis of unitarity considerations). 
This creates the association ``imaginary potential''
$\rightarrow$ ``desappearance of the flux''.    
For a black sphere potential, following 
the values of the high energy wavefunction along a 
line with impact parameter $\vec b$ we have the total 
damping $\psi_{exit}/\psi_{entrance}$ $=$ 
$exp(-mW\sqrt{R^2-b^2}/k)$. 
So in quite a natural way the matching conditions 
oblige the external wavefunction to be large 
at the entrance side, and small on the exit side.  
Saturation of the bulk absorption appears 
for $R$ $>>$ $k/mW$: then the internal function 
is $\approx$ 0 on the exit side whatever $R$, and we 
speak of surface absorption with $\sigma$ $\sim$ $R^2$.

In practical problems, 
we can $sometimes$ speak of ``high energy'' down to a 
situation where only the S and P waves are important.
In such a case the interference between the two waves 
must be effective 
enough to assure a relevant difference between 
$\vert \psi_{exit}\vert^2$ and 
$\vert \psi_{entrance}\vert^2$. In $\bar{p}-$nucleus 
the high energy approximations, as described in the 
introduction, can be pushed with some 
residual effectiveness down to momenta $\sim$ 100 MeV/c. 

In the case of the low energy 
S-wave dominated problem (with the same 
black sphere target) there are relevant differences both 
in the behavior of the internal wave and  
in the use of the matching conditions. First, there 
are no more 
possibilities of reducing the Schroedinger equation 
to a first order one. The solution becomes 
practically 1-dimensional with respect to the radius: 
$\psi(\vec r)$ $\approx$ $\psi(r)$. 
In the limit $k$ $\approx$ 0 
(or reasonably: whenever $k^2/2m$ $<<$ $W$) 
the problem 
admits two solutions of the kind $exp(\pm \sqrt{-2imW}r)/r$
$\sim$ $exp(\pm [(1-i)\sqrt{mW}r]/r$ 
(from now on we will discuss of the 
behavior of $r\psi$ instead of $\psi$, so to avoid 
the accompanying factor $1/r$).  
As in any scattering problem, the overall normalization 
of the wavefunction is not relevant (the observables 
are determined by the ratios between incoming and outgoing 
fluxes). The asymptotic form of the external (S-wave)
wavefunction must be of the form 
$r\psi$ $=$ $g sin[k(r-a)]$, 
with constant $g$ and $a$ ($a$ is the scattering length). 
This function is the sum of an incoming and an outgoing 
wave $exp[\pm ik(r-a)]$. A nonzero imaginary part of 
$a$ means net flux absorption in the scattering, 
i.e. inelastic processes. The total elastic and inelastic 
parts of the cross section are proportional 
to $\vert a\vert^2$ and to 
$\vert Im(a)\vert/k$ (see e.g. \cite{ll1}). On this 
ground one can say that at low energies 
the ``geometric'' naive expectations for the values of the 
cross sections are $\sigma_{el}$ $\sim$ $R^2$ and 
$\sigma_{react}$ $\sim$ $R/k$, reached when $Re(a)$ $\sim$
$Im(a)$ $\sim$ $R$. 

The two boundary conditions to be satisfied 
are: (1) Finiteness 
in the center, which obliges one to consider both 
the above solutions 
with relative coefficient $-1$:
$r\psi$ $\sim$ 
$exp([(1-i) \sqrt{mW}r]- exp(-[(1-i) \sqrt{mW}r]$. 
(2) Matching 
with the total external solution in $r$ $=$ $R$ (the 
target surface). Because of the lack of meaning of the 
overall normalization of the solution, the matching 
condition is normally required for the logarithmic 
derivative of the solution: 
$[\psi'/\psi]_{internal}$ $=$ 
$[\psi'/\psi]_{external}$ 
(at $r$ $=$ $R$); $\psi_{external}$ is the sum of the 
inward and outward directed spherical waves. 

We stress this point: 
The $S-$wave low energy problem only knows 
$one$ matching place, 
``entrance'' and ``exit'' now coincide. 
In their 1-dimensional forms (with respect to $z$ and 
$r$ variables respectively)  
the high energy 
problem is a ``transition through a wall'' problem,
while the low energy one is a ``reflection'' 
problem. 

As we have noticed above, 
now one can't discard by assumption any of the 
two components 
$exp(\pm [(1-i) \sqrt{mW}r]/r$ 
of the internal solution. They are both necessary 
to satisfy the condition of finiteness in the center. 
Since their relative coefficient is $-1$, if $R$ $>>$ 
$1/\sqrt{mW}$ one of the two terms only is relevant 
in $r$ $=$ $R$. This is the saturation of bulk absorption 
in the low energy problem. Intuition suggests that 
for any $R$ large enough to satisfy the saturation 
condition $R$ $>>$ $1/\sqrt{mW}$ the full spherical 
incoming wave disappears inside the sphere, and it is 
not possible to remove more than the full flux. 
In semiclassical regime, this would imply a large 
flux absorption. This is not the case here. 
It may be verified, by direct calculation, that  
when the internal 
function at the matching point is composed by only one 
of the above two terms, 
matching can be fulfilled only if the inward and 
outward directed external waves have the same 
magnitude. 
 
More insight in the black sphere subtilties can be reached 
by a comparison with a related 
problem: the scattering by a (very) hard core elastic 
repulsive potential $V(r)$ $=$ $+W$ for $r$ $<$ $R$,
with $W$ $>>$ $k^2/2m$.   
In both problems the modulus of the internal 
solution behaves the same way, i.e. damps exponentially 
to zero on a scale 
$R-r$ $\sim$ $1/\sqrt{mW}$, and 
the logarithmic derivative (in the matching point) 
of the internal wavefunction is very large for strong
potentials, has positive real part, and imaginary part 
not larger in magnitude than the real part (in the 
hard core case of course the imaginary part is zero). 
This has the consequence that 
the scattering length is very close to $R$. We remind 
the geometrical meaning of the scattering length: 
At very small 
$k$ and in the surroundings of the matching point $r$ $=$ 
$R$, the $external$ function $r\psi$ can be written 
in the form $r\psi$ $=$ $(B/k) sin [k(r-a)]$ and 
approximated by the straight line $B (r-a)$.  
The constant $a$ is the scattering length, and 
for $r$ $=$ $a$ 
the virtual prolongation of the linearized external 
solution will reach zero (of course the internal real 
solution will not be zero at $r$ $=$ $a$, usually). 
The practical key of 
nonrelativistic scattering problems  
is of course the possibility to calculate 
somehow the internal solution. We suppose one has been 
able to do it, and has normalized the internal 
wave in some arbitrary way. 
By applying the matching condition, i.e. by requiring 
the equality of the external and internal logarithmic 
derivatives, one determines 
all the parameters of the external wave, in particular 
$a$. 
If the logarithmic derivative is large enough 
($\psi'/\psi$ $>>$ $1/R$) the virtual 
prolongation of the external solution will reach zero 
for $r$ very close to $R$, so $a$ $\approx$ $R$. 
This is what happens with a very large real repulsive
potential. Incidentally we note that the 
S-wave boundary condition for the $free$ 
motion wavefunction (i.e. $r\psi_{free}(r)$ $=$ $sin(kr)$ 
$=$ 0 for $r$ $=$ 0) defines the origin as the ``source''
of the outgoing waves. In presence of a repulsive potential
the position of this source (which is now virtual) 
is ``moved'' to some point 
$r$ $=$ $a$ $>$ 0. For a repulsive potential of 
infinite strength, $a$ $=$ $R$. 
In the case of a black sphere potential in saturation 
condition, the reported form 
of the internal solution suggests that the real and 
imaginary 
part of the logarithmic derivative are of the same scale at 
the matching radius. Then a large logarithmic derivative  
implies 
$Re(a)$ $\approx$ $R$,  
$Im(a)$ $\sim$ $R-Re(a)$  
(the second equation would not be 
necessarily valid if the imaginary 
part of the logarithmic derivative were much larger than 
the real part; this corresponds to the case of a potential  
dominated by a negative real part). 
In the limit of a very large $W$, 
$Re(a)$ $\rightarrow$ $R$,   
$Im(a)$ $\rightarrow$ 0, 
meaning 
an elastic cross section close to the geometrical section,
and reaction cross section $<<$ $R/k$, i.e. much smaller 
than what one could expect assuming a size $R$ of the absorbing 
region. An approximate calculation, valid for $\sqrt{mW}$ $>>$ 
$k$ and $\sqrt{mW}$ $>>$ $1/R$, 
gives $R-a$ $\approx$ $0.5(1+i)/\sqrt{mW}$. 
So, when the two conditions $mWR^2$ $>>$ 1 and 
$W$ $>>$ $k^2/2m$ are realized, the 
black sphere and the hard core sphere lead to similar 
results, and we find that a potentially huge internal 
absorption just implies very good reflection of the 
incoming flux. 

There is an interesting difference 
between the two problems: in the hard core problem, the internal 
fastly damped wave is a stationary wave, with no net flux.
In the black sphere problem a flux $\propto$ $-exp(2\sqrt{mW}r)$ 
(i.e. inward directed) is present inside the sphere. 
Despite this, real flux absorption is small for large $W$. 
This does not create contraddictions. In fact, crossing 
the matching point $d\psi/dr$ is continuous and regular, 
implying that the velocity is more or less the same on both 
sides, in the surroundings of $r$ $=$ $R$. 
However, due to 
the large and positive $Re(\psi'/\psi)$ in $r$ $=$ $R$,
and to the smallness of $k$, 
the overall normalization constant $b$ of the external 
solution $b sin[k(r-a)]$ is much larger than any value 
of $\psi$ inside the potential region. For this reason, 
all the flux that is able to cross the potential surface 
is definitely lost, but this is much 
smaller than the incoming flux. Notice that, for large $k$,  
a large $\psi'/\psi$ in the matching point would not imply 
a large value of $b$. 

We stress that in the above discussion the key points 
leading to a small flux removal are: 
(1) entering a well defined space region (the nucleus)
the modulus of the projectile wavefunction 
becomes fastly much smaller than 
in the external space. More precisely we need 
$\vert \psi'/\psi\vert $ $>>$ $1/R$ $>>$ $k$; 
(2) the imaginary part of 
the logarithmic derivative of the wavefunction 
at the target surface is not much larger in scale 
than the real part. As far as one believes that these 
two conditions are respected in general in 
the $\bar{p}-$nucleus annihilation problem, the 
potential can be completely forgotten and the above 
conclusions on inversion, saturation etc. remain valid. 
In other words, 
even though one does not believe to the reliability of 
an optical potential model for $\bar{p}-$nucleus 
interactions, the fact that a strong absorption of 
the projectile wave inside the target 
can be associated with an almost complete flux 
reflection outside seems to be completely general: 
it originates 
in the regularity properties of the projectile wavefunction 
at the surface of a strong absorber 
and not in the details of the absorption mechanism. 
This is in agreement with 
the final words of ref.\cite{wic}, where the ``repulsive 
properties of absorption'' are explicitely cited to 
explain the opposite effect of single and double 
scattering terms. 

Concerning magnitudes, 
for a target radius of 2 fm the saturation condition 
is reached for $W$ $\sim$ 10 MeV, and 
larger imaginary potentials 
or larger radii don't imply more annihilations. 
The above discussed loss of semiclassical intuition in 
this problem prevents us from attributing a precise 
meaning to a given 
value of the imaginary part of an optical potential. 
However, intuition can content itself with the fact 
that the value of $W$ is associated with the spatial 
scale $1/\sqrt{mW}$ over which we assume the external 
flux to have been reasonably damped inside the nucleus. 
In this sense $1/\sqrt{mW}$ can be interpretated 
as the ``free mean 
annihilation path in nuclear matter'', but this 
number has nothing to do with the rate of annihilation 
we really see. In the limit where this path tends to 
zero we see no annihilations at all. 

The consequence of the previous discussion is that in 
$systematic$ analysis of $\bar{n}$ annihilations 
on large-$A$ nuclei at low energies one should expect   
a roughly constant 
cross section. With $\bar{p}$
projectiles the cross section would be proportional to 
the nuclear charge $Z$. 
Behind this there is the assumption that small portions 
of hadronic matter (i.e. a single nucleon) have very large 
absorption properties. 

With very light nuclei, structure details 
can be more important than nuclear matter 
bulk properties (e.g. 
one can't speak of deep nucleons which are 
untouched by annihilation processes). 
Our analysis 
of the damping properties of the wavefunction relies 
on target continuity properties which could be 
absent in this case. Taking into account that Coulomb 
forces are not able, in very light nuclei, 
to accelerate $\bar{p}$ up to Fermi motion 
typical velocities, 
we could imagine two opposite limiting pictures. 
In the first one, before being subject 
to hard (elastic or not) nuclear interactions the 
antinucleon has no time to undergo relevant soft 
interactions, that would accelerate it to 
momenta $k$ $\sim$ 
$p_{Fermi}$.  
Then, due to the difference in velocity, 
we can speak of approximate adiabaticity, and say that 
$\bar{p}$ 
sees the nucleus more or less as a continuous target. 
In such a case the black sphere results  
could be applied to light nuclei too 
(very approximately, since the difference in the 
velocity scale is not huge in the actual experimental 
kinematics\cite{obe1,obe2}). The opposite picture is 
the ``compound nucleus'' scenario: 
The antinucleon is able to mix with the target 
nucleons forming a ``compound nucleus'', where all 
hadrons share similar velocities. In this case some 
nuclear-scale time 
would pass between the disappearance of the elastic 
channel and the formation of any final state, with 
presence of narrow resonances and large fragmentation 
probability 
for the residual nucleus.  
Presently we don't see (or don't resolve) resonances 
or many fragment final states 
(such phenomena could be present already at projectile 
momenta $\sim$ $p_{Fermi}$ $\approx$ 200 MeV/c). 
If the compound nucleus picture were realistic 
the above presented arguments should be largely 
modified. 

Generalizing the last argument, if the $\bar{p}-$nucleon 
interaction is characterized also by a strong 
(elastic) attracting part, with the same range 
as the nucleon-nucleon interaction, 
this could result in the formation 
of a global nuclear attracting well, 
with radius $\sim$ the nuclear radius and diffuseness
$1/m_\pi$. Such an attracting well could produce 
narrow ($\bar{p}-$nucleus) resonances. 
As the potential 
model itself can show, by just taking a general 
potential with form $V(r)$ $=$ $-(U+iW)$ and studying 
numerically some random solutions, the presence 
of resonances 
may largely affect the validity of all the previous 
conclusions. In particular these solutions may present 
``phantom'' resonances. 
With this term we indicate cross section peaks 
which would 
be clear and narrow for $W$ $=$ 0, and disappear with 
a nonzero $W$ $\sim$ $U$ (which should be the 
case in our problem). Phantom resonances would not 
be easily detected, but enhance anyway the 
$\bar{p}-$nucleus annihilation cross section. 
Qualitatively the effect of a resonance 
can be understood by observing 
that any attractive interaction will act as a 
focusing device (as it happens with the Coulomb 
interaction), but in particular a 
$narrow$ resonance may enlarge much the time 
spent by the projectile in the interaction region. 
In the previous language: it will enlarge the 
value of $\vert \psi\vert^2$ inside the target,
so decreasing the value of the logarithmic derivative 
at the surface. This may increase largely the  
annihilation rate, and a systematical analysis of the 
phenomenon does not appear easy to us. For the case of 
a Deuteron target the problem is under examination in 
\cite{pro2}. 

In the same work\cite{pro2} it has been noted 
that the behavior 
of the black potential sphere can be largely affected 
by the presence of a long exponential 
tail of the potential shape 
(we have assumed, in the previous discussion, 
a sharp potential surface). In the terms used in the 
previous discussion, where the potential is strong 
the $\bar{p}$ wavefunction $\psi$ 
will be fastly damped to zero. This creates a 
large relative 
variation of $\psi$ in the interaction surface region.  
But if this region diffuses over a region 
of size $\sim$ $R$, $\psi$ will pass from its external 
value to zero in a region of size $R$, and the previous 
requirement $\psi'/\psi$ $>>$ $1/R$ will not be realized. 
To see it in anothwer way: the ``black'' sphere is 
surrounded by a large ``grey'' cloud, were annihilation 
may be much more effective (in agreement with the above 
established law: smaller potential $\rightarrow$ 
stronger annihilation rate). 
The condition of a diffuseness $\sim$ $R$ will not 
be realized with heavy nuclei, but is possible with small 
compact nuclei like $^4$He, 
were $R$ is not much larger than 
$1/m_{\pi}$, the reasonable upper limit for any 
strong interaction diffuseness. 

Concluding this section, on 
one side the study of the wavefunction suggests 
on a very general ground that a 
saturation behavior should be present and detectable 
in $\bar{p}$ and $\bar{n}$ annihilation on large-A nuclei, 
but on the other side 
it is still doubtful that this may be considered $the$  
satisfactory explaination of 
the inversion behavior that we actually 
see in very light nuclei. We must even remember that 
this saturation behavior means a systematic $\sim$ $Z$ law
for $\bar{p}-$nucleus annihilations. So the saturation 
could (roughly) be enough to  
explain the ratio between $\bar{p}-p$ and $\bar{p}-D$ data, 
but surely not the data on $^4$He. 

\section{Summary and conclusions}

This work has been motivated by the need of understanding 
the apparently anomalous behavior of $\bar{p}-$nucleus 
annihilation cross sections at low energies. 

Most of our efforts have been devoted to an 
analysis of the impulse 
approximation contribution to the cross section 
on nuclei. In general terms, and taking Coulomb effects into 
account, this has been expressed by 
eq.(\ref{free12}), with eq.(\ref{free11}) as reasonable 
approximation for momenta over 100$\div$200 MeV/c. 

For the limit of momenta below 100 MeV/c the 
ratio $\equiv Z R_A$ between the nuclear and single nucleon 
annihilation cross sections has been extimated 
via eq.(\ref{free17}). 
This ratio, which has been calculated in an approximated way 
via an average on 
the squared modulus of the matrix element for the low energy 
$\bar{p}-$nucleon annihilation, 
has the peculiarity of depending on nuclear quantities 
only. Since, however, 
a complete calculation of eq.(\ref{free17}) 
requires a complicated numerical Montecarlo integration, 
and knowledge of the nuclear momentum distribution, 
we have just performed some estimation for the special 
case of two 
pion emission, within mean field nuclear distributions. 
These estimations simply reproduce the most expected 
PWIA behavior of cross sections, i.e. $A$ times the 
elementary cross sections ($AZ$ when Coulomb corrections 
are included). 
For momentum distributions which are large at 
large momenta (meaning the presence of strong non mean 
field contributions) we find a decrease with respect to 
the $\sim A$ expectation, although the effect is not 
very marked (a 10 \% correction, in magnitude). 

PWIA allowed us to simulate the 
characteristical Fermi motion smearing of a resonance 
peak (fig.5). The 
conclusion is that $in$ $principle$ this mechanism is 
able to produce 
inversion exactly the way we detect it (it is just 
a matter of tuning parameters),  
however some possible, but rather restrictive,  
conditions should be fulfilled. In particular, resonances 
in the $\bar{p}-$nucleon cross section should be 
very narrow and all 
regrouped within a small momentum range just a little 
below the lower limit of the experimentally analyzed 
region. 

Accepting that 
IA (in practics: single scattering processes) 
is unable to 
explain both the experimental data and the 
potential model strangenesses, we feel like 
supporting the role of 
multiple scattering effects, as somehow indicated in 
\cite{pro1} and \cite{wic}. We have discussed 
the possible physical origin of such effect, 
indicating  
several among the possible 
mechanisms (elastic shadowing, 
inelastic shadowing and antishadiwing, and soft distortions 
of the $\bar{p}$ wavefunction) 
which would probably require 
a separate style of analysis each. Even in the 
relatively simple case of a Deuteron target, an exaustive  
calculation of the effect of inelastic 
multiple scattering looks presently prohibitive. 

While discussing Coulomb interactions, we have 
perfectly reproduced the $\bar{p}p$ annihilation cross 
section with a very simple optical model 
(a pure imaginary potential well with Woods-Saxon 
shape, and standard parameters for radius and 
diffuseness). Taking into 
account that the optical model  
is also able to produce inversion,  
we have devoted a section 
of this work to an extensive analysis of the behavior 
of the solutions of the optical model in its simplest form,
i.e. the black sphere. The conclusion is 
more general than this model would allow for: 
it appears that whenever in a spherical region of space 
the projectile wavefunction is heavily suppressed 
almost all 
of the incoming flux must be elastically reflected. 
This is an obvious result when the suppression mechanism 
is elastic repulsion, but we find that the same must 
happen even when the mechanism which creates 
a ``vacuum hole'' in the projectile flux is inelastic. 
This phenomenon looks more suitable for application  
to the case of heavy nuclear targets  
(where the general nuclear matter 
properties are more relevant than the details 
of the structure) than for the case of 
the light nuclei used in the performed experiment\cite{obe1}.   
In addition it 
would anyway suggest a value of the 
$\bar{p}-^4$He annihilation rate 
larger than the $\bar{p}-p$ one (because of the 
stronger Coulomb attraction) 
instead of the observed inversion between the two. 

So, although some explainations are 
available for the detected inversion mechanism, 
the problem is still open.


\newpage

\end{document}